\documentclass[a4paper]{report}
\usepackage[utf8]{inputenc}
\usepackage[T1]{fontenc}
\usepackage{RJournal}
\usepackage{amsmath,amssymb,array}
\usepackage{booktabs}

\usepackage{orcidlink,thumbpdf,lmodern}
\usepackage{framed}
\usepackage{amsmath}
\usepackage{amssymb}
\usepackage{multirow}
\usepackage{booktabs}
\usepackage{makecell}
\usepackage{seqsplit}
\usepackage{mathtools}
\usepackage{amsthm}
\usepackage{array,epsfig,fancyheadings,rotating}
\usepackage{amsfonts}
\usepackage{tabularx}

\DeclarePairedDelimiter\floor{\lfloor}{\rfloor}

\theoremstyle{definition}
\newtheorem{exmp}{Example}[section]
\usepackage{titlesec} 
\titleformat{\subsection}[hang]
  {\normalfont\fontsize{10.5}{12}\bfseries}{\thesubsection}{1em}{}
\renewcommand{\thesubsection}{\arabic{section}.\arabic{subsection}}
\renewcommand\thesection{\arabic{section}}
\newcolumntype{b}{X}
\newcolumntype{s}{>{\hsize=.1\hsize}X}
\newcolumntype{m}{>{\hsize=.3\hsize}X}

\begin{document}

\sectionhead{Contributed research article}
\volume{XX}
\volnumber{YY}
\year{20ZZ}
\month{AAAA}

\begin{article}
\title{CDsampling: An R Package for Constrained D-Optimal Sampling in Paid Research Studies}
\author{by Yifei Huang, Liping Tong, and Jie Yang}

\maketitle

\abstract{
In the context of paid research studies and clinical trials, budget considerations often require patient sampling from available populations which comes with inherent constraints. We introduce the R package CDsampling,  which is the first to our knowledge to integrate optimal design theories within the framework of constrained sampling. This package offers the possibility to find both D-optimal approximate and exact allocations for samplings with or without constraints. Additionally, it provides functions to find constrained uniform sampling as a robust sampling strategy when the model information is limited. To demonstrate its efficacy, we provide simulated examples and a real-data example with datasets embedded in the package and compare them with classical sampling methods. Furthermore, the CDsampling package revisits the theoretical results of the Fisher information matrix for generalized linear models (including regular linear regression model) and multinomial logistic models, offering functions for its computation.
}

\section{Introduction}\label{sec:intro}

Paid research studies are essential for determining the influence of new interventions or treatments for providing quantitative evidence in various domains, such as healthcare, psychology, and politics. However,  conducting such studies often involves a limited budget and a large pool of potential volunteers, which poses a challenge in selecting the best sample to meet the research objectives. Poor samplings may result in biased or inaccurate estimates, low statistical powers, and even misleading conclusions. Therefore, finding a good sampling strategy is crucial for researchers and practitioners. 

Consider a constrained sampling problem commonly encountered in paid research studies. Suppose, in a research study to evaluate a new intervention, $N=500$ volunteers register to participate. Upon registration, the investigators collect basic demographic information, such as gender (female or male) and age groups ($18\sim25$, $26\sim64$, $65$ and above) for each volunteer.  Treating gender and age as stratification factors, we obtain $m=6$ subgroups. However, due to budget limitations, the study could only accommodate $n=200$ participants. Let $N_i$ denote the frequency of volunteers within the $i$th subgroup, where $i=1,\dots,m$. We call the integer number of participants sampled from each subgroup, $n_i$, as the exact allocation, and the corresponding proportion $n_i/N_i$ as the approximate allocation. The goal is to select a sample of $200$ participants $\mathbf n = (n_1, \dots, n_m)$, such that $\sum_{i=1}^m n_i = 200$ from the pool of $N=500$ volunteers to evaluate the intervention effect most accurately, subject to the constraint that no subgroup is oversampled beyond the number of available volunteers within that subgroup, that is, $n_i \le N_i$. This constraint is the most commonly encountered in paid research studies (see Section~\ref{sec:example_glm} for details, and for constraints of other forms, please refer to Section~\ref{sec:MLM_example}).

Commonly used sampling strategies include simple random sampling, stratified sampling, and cluster sampling. Simple random sampling is the most straightforward form of probability sampling, where each element has an equal chance of being selected \citep{tille2006sampling}. Proportionally stratified sampling involves dividing the population into homogeneous subgroups, such as gender and age groups, and applying random sampling to sample proportionally within each subgroup \citep{lohr2019sampling}. Cluster sampling, on the other hand,  splits the population into heterogeneous clusters, for example, based on the locations of volunteers, and randomly selects some clusters as the sample units. However, these methods have their drawbacks. Cluster sampling is relatively low-cost but less precise than simple random sampling. Stratified sampling can produce more efficient estimators of population means, but it requires finding well-defined and relevant subgroups that cover the entire population \citep{alma991096987459705816}. Moreover, these existing methods are based on assumptions that may not hold if the model estimation is the primary goal of the paid research study.

In the proposed \CRANpkg{CDsampling} package, we implement the sampling strategy based on optimal design theory (with main functions \texttt{liftone\_GLM()}, \texttt{liftone\_constrained\_GLM()}, \texttt{liftone\_MLM()},  and \texttt{\seqsplit{liftone\_constrained\_MLM()}}), which can improve the accuracy of the intervention effect estimation. Optimal design theory is a branch of experimental design that aims to find the best allocation of experimental units to achieve a specific optimality criterion such as minimizing the variance of estimation or equivalently maximizing the information obtained from the design. For example, D-optimality maximizes the determinant of the information matrix, which minimizes the estimators' expected volume of the joint confidence ellipsoid. A-optimality minimizes the trace of the inverse information matrix, equivalent to minimizing the average of the variances of the estimators. E-optimality minimizes the maximum eigenvalue of the inverse information matrix, which minimizes the largest expected semi-axis of the confidence ellipsoid and protects against the worst possible case \citep{fedorov1972,atkinson2007,yangmin2012,fedorov2014}. In this paper, we focus on D-optimality due to its overall good performance and mathematical simplicity \citep{atkinson1999,atkinson2007}. \textcolor{black}{According to \cite{huang2023constrained}, the constrained D-optimal sampling strategies work well for paid research studies or clinical trials.} To implement the recommended sampling strategy, we develop an R package called \CRANpkg{CDsampling} (namely, \underline{C}onstrained \underline{D}-optimal \underline{sampling}), available on CRAN at \url{https://cran.r-project.org/package=CDsampling}. To our best knowledge, the \CRANpkg{CDsampling} is the first R package offering the sampling tool with constraints in paid research studies based on optimal design theory. Package \CRANpkg{sampling} implements random samples using different sampling schemes \citep{tille2016package}. The package also provides functions to obtain (generalized) calibration weights, different estimators, as well as some variance estimators. Package \CRANpkg{SamplingStrata} determines the best stratification for a population frame that ensures the minimum sample cost with precision thresholds \citep{barcaroli2014samplingstrata}. On the other hand, there are existing R packages for optimal designs. Package \CRANpkg{AlgDesign} finds exact and approximate allocations of optimal designs under D-, A-, and I-criteria \citep{wheeler2019package}.  Package \CRANpkg{OptimalDesign} enables users to compute D-, A-, I-, and c-efficient designs with or without replications, restricted design spaces, and under multiple linear constraints \citep{harman2016package}. Package \CRANpkg{acebayes} find Bayesian optimal experimental designs with approximate coordinate exchange algorithm \citep{overstall2017acebayes,overstall2018package}. Package \CRANpkg{OPDOE} provides functions for optimal designs with polynomial regression models and ANOVA \citep{gromping2011optimal}. These packages provide programming tools for finding samplings or optimal designs under different criteria and models. However, they do not address the specific challenges of practical feasibility in the constrained sampling scheme of paid research studies. The proposed \CRANpkg{CDsampling} package fills this gap by offering an efficient sampling tool to handle general constraints and common parametric models in paid research studies.

\section{Method}\label{sec:model}
\subsection{Constrained lift-one algorithm}\label{sec:model_algorithm}
The lift-one algorithm was initially proposed by \citet{ymm2016} to find D-optimal designs with binary responses. This was extended to generalized linear models (GLMs) by \citet{ym2015} and subsequently adapted for cumulative link models by \citet{ytm2016}. The methodology was further extended to multinomial logit models (MLMs) by \citet{bu2020}. 

\begin{figure}[htb!]
    \centering
    \hbox{\hspace*{-1em}\includegraphics[width=1.1\textwidth]{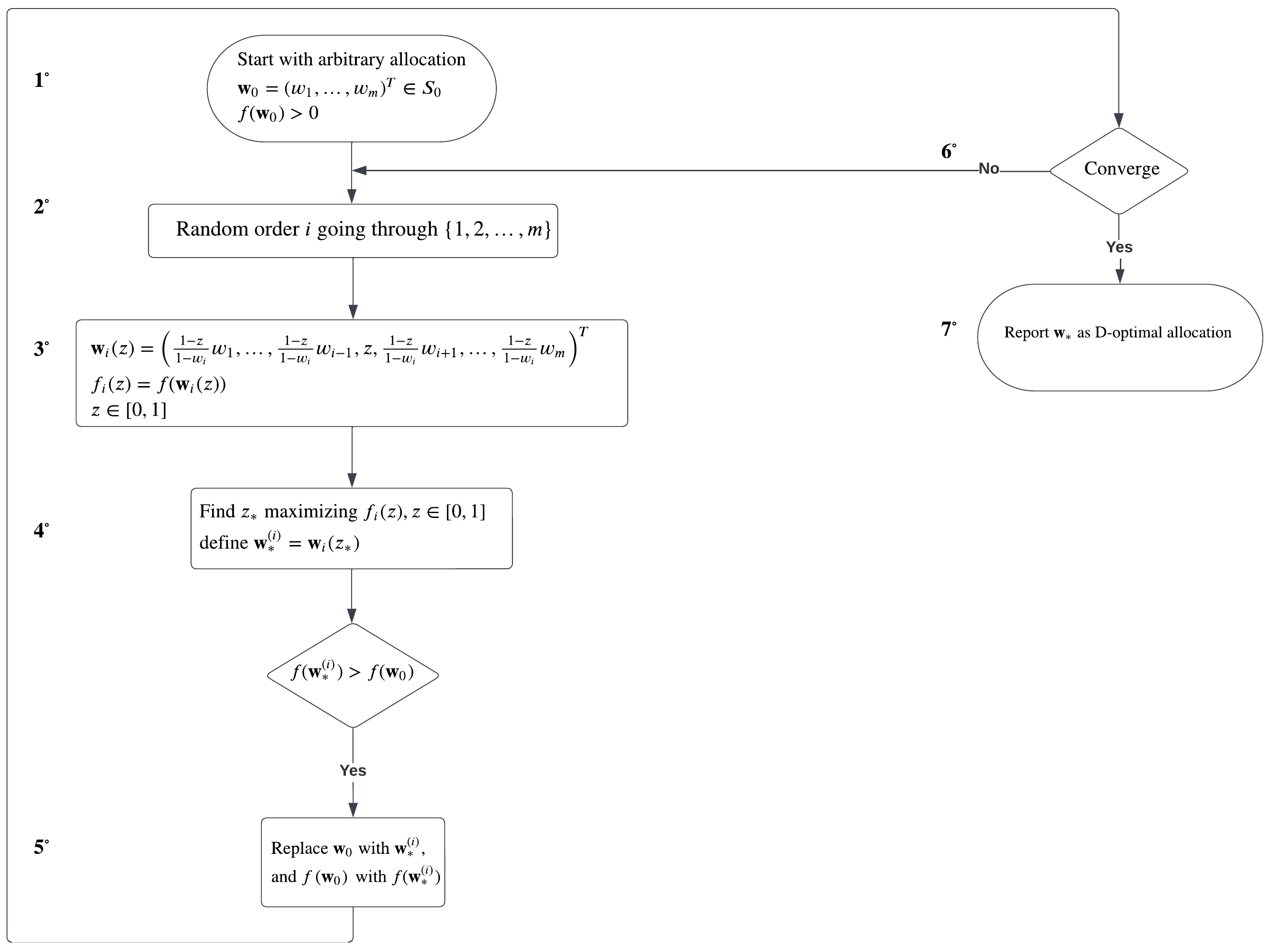}}
    \caption{The framework of lift-one algorithm.}
    \label{fig:liftone_algo}
\end{figure}

Figure~\ref{fig:liftone_algo} provides a concise summary of the lift-one algorithm applied to general parametric models. The detailed algorithm is provided in Algorithm~$3$ from the Supplementary Material of \citet{huang2023constrained}. We consider a general study or experiment involving covariates $\mathbf{x}_i = (x_{i1}, \dots, x_{id})^\top$, for $i = 1, \dots, m$, referred to as experimental settings. In paid research studies, these covariates could be the stratification factors such as the gender and age groups in our motivating example. Here, $m \ge 2$ denotes the number of experimental settings, which corresponds to $m=6$, the number of gender and age groups in the motivating example. Suppose the responses follow a parametric model $M(\mathbf x_i, \boldsymbol \theta)$, where $\boldsymbol \theta \subseteq \mathbb{R}^p$ with $p\ge 2$. Under regularity conditions, the Fisher information matrix of the experiment design can be written as $\mathbf F = \sum_{i=1}^m n_i \mathbf F_i$, where $\mathbf F_i, i=1, \dots, m$ is the Fisher information matrix at $\mathbf x_i$ and $n_i$ is the number of subjects allocated to the $i$th experimental setting. In the setting of paid research studies, $n_i$ corresponds to the number of subjects sampled from the $i$th subgroup. We usually call $\mathbf n = (n_1, \dots, n_m)^\top$ the exact allocation, where $n=\sum_{i=1}^m n_i$, and $\mathbf w = (w_1,\dots,w_m)^\top=(n_1/n, \dots, n_m/n)^\top$ the approximate allocation, where $w_i \ge 0$ and $\sum_{i=1}^m w_i = 1$ \citep{kiefer1974, pukelsheim2006optimal, atkinson2007}. The approximate allocation is theoretically more tractable while the exact allocation is practically more useful. In the statistical literature, approximate allocations are more commonly discussed \citep{kiefer1974}. The D-optimal design aims to find the optimal allocation that maximizes $f({\mathbf w})=|\mathbf F({\mathbf w})|=|\sum_{i=1}^m w_i \mathbf F_i|$. The lift-one algorithm simplifies a complex multivariate optimization problem into a sequence of simpler univariate optimization problems.  This is achieved by ``lifting'' and optimizing one weight $w_i$, within the approximate allocation vector $\mathbf w$. Specifically, in step $3^\circ$ of the algorithm depicted in Figure~\ref{fig:liftone_algo}, the determinant of the Fisher information matrix function concerning the lift-one variable is expressed as $f(z) = f(\mathbf w_i(z))$ where the variable $z$ substitutes for $w_i$ in allocation $\mathbf{w}$, and the remaining weights are adjusted proportionally, denoted as $\mathbf w_i$. The updated weight vector is given by 
\[
\mathbf w_i(z) = \left( \frac{1-z}{1-w_i}w_1, \ldots, \frac{1-z}{1-w_i}w_{i-1}, z, \frac{1-z}{1-w_i}w_{i+1}, \ldots, \frac{1-z}{1-w_i}w_m\right)^\top.
\]
The allocation that results from the convergence in step $6^\circ$ of the algorithm is identified as the D-optimal approximate allocation.

\begin{figure}[htb!]
    \hbox{\hspace*{-3em}\includegraphics[width=1.18\textwidth]{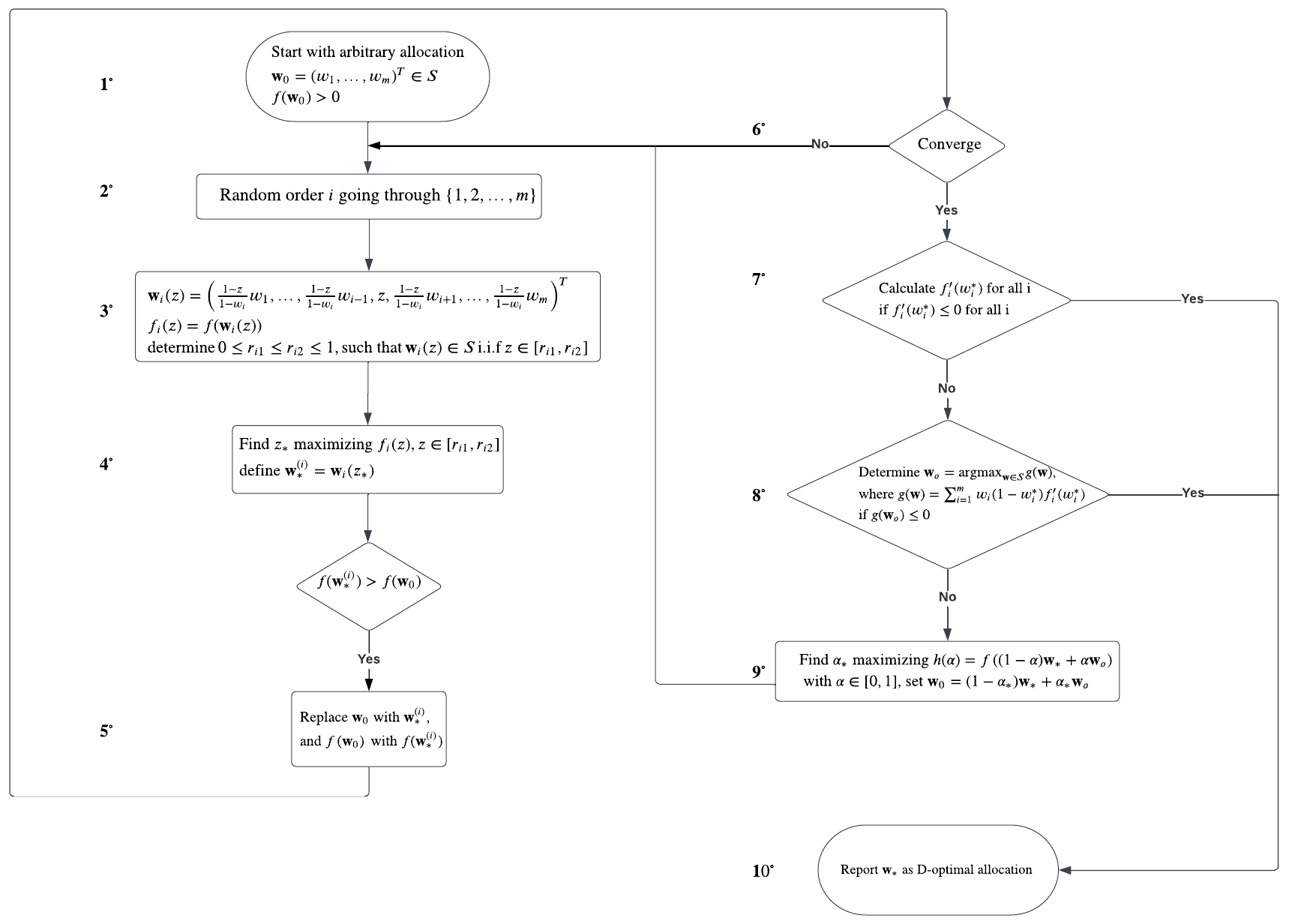}}
    \caption{The framework of constrained lift-one algorithm.}
    \label{fig:constrained_liftone_algo}
\end{figure}

In the context of paid research studies, budgetary limitations often necessitate the selection of a subset of participants. We consider the sampling of $n$ subjects from a larger population of $N$, where $n < N$. A typical constraint in such studies is $n_i \leq N_i$, where $N_i$ represents the number of available subjects from the $i$th experimental setting group. This effectively places an upper bound on the sample size for each subgroup (or stratum) ensuring the sampling process does not overdraw from any subgroup. Additional constraints may include $n_1 + n_2 + n_3 + n_4 \leq 392$ (see the MLM example in Section~\ref{sec:MLM_example}), $4n_1 \geq n_3$ (see Example~\ref{ex:simple_logistic_r1_r2} in Supplementary Material  Section~\ref{sec:supp_comparison}), etc. The constrained lift-one algorithm seeks the approximate allocation ${\mathbf w}$ that maximizes $|{\mathbf F}({\mathbf w})|$, on a collection of feasible approximate allocations $S \subseteq S_0$ where $$S_0 := \{(w_1, \ldots, w_m)^\top \in \mathbb{R}^m \mid w_i \geq 0, i=1, \ldots, m; \sum_{i=1}^m w_i = 1\}.$$ The set $S$ is presumed to be either a closed convex set or a finite union of closed convex sets.  The framework of the constrained lift-one algorithm is provided in Figure~\ref{fig:constrained_liftone_algo}. The details of the algorithm can be found in Algorithm~1 of \citet{huang2023constrained}. In the constrained lift-one algorithm, the search for the optimal lift-one weight $z$ in step $3^\circ$ of Figure~\ref{fig:constrained_liftone_algo} is confined within the interval of $[r_{i1}, r_{i2}]$ (see Example~\ref{ex:simple_logistic_r1_r2} in Supplementary Material and Section S.3 in \cite{huang2023constrained} for details of finding $[r_{i1}, r_{i2}]$).   To ensure that the resulting allocation is D-optimal, two additional decision steps, labeled steps $7^\circ$ and $8^\circ$, are incorporated into the algorithm. The reported allocation in step $10^\circ$ is the constrained D-optimal approximate allocation for the study. To illustrate the difference between the lift-one algorithm and the constrained lift-one algorithm, we provide examples in Supplementary Material Section~\ref{sec:supp_comparison}. 

\begin{figure}[htb!]
    \centering
    \includegraphics[width=0.95\textwidth]{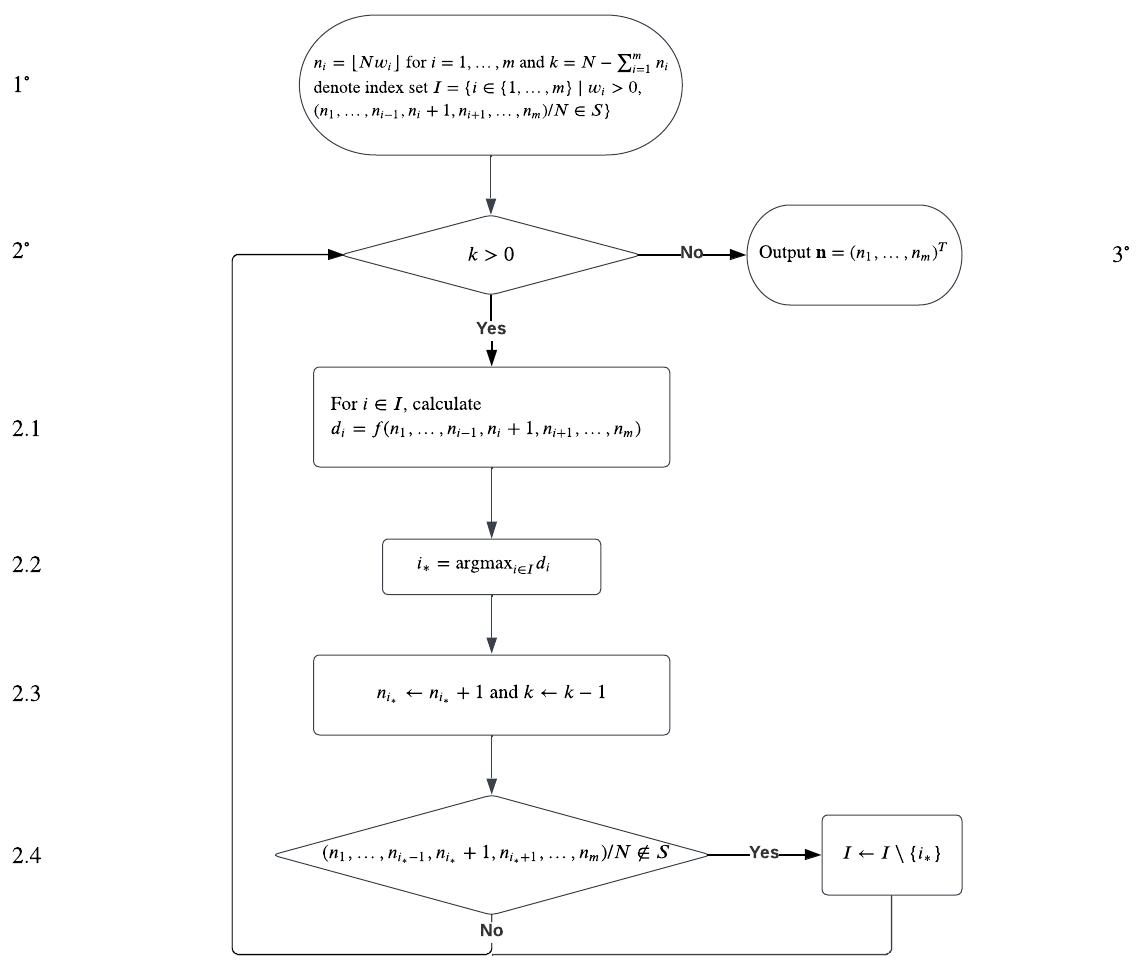}
    \caption{The framework of constrained approximate to exact algorithm.}
    \label{fig:approx_to_exact_algo}
\end{figure}

Upon obtaining the approximate allocation, we may employ the constrained approximate to exact allocation algorithm outlined in Figure~\ref{fig:approx_to_exact_algo} for the conversion of a real-valued approximate allocation to an integer-valued exact allocation. The full details of the algorithm are in Algorithm~2 of \citet{huang2023constrained}. The algorithm begins by assigning a floor integer value to all subgroups $n_i = \floor{Nw_i}$. Subsequently, each remaining subject is added to the corresponding group in a manner that maximizes the determinant of the Fisher information matrix. This transformation provides a more pragmatic application in the actual sampling process within paid research studies.

The calculation of the Fisher information matrix $\mathbf{F}$ and the subsequent maximization of its determinant $|\mathbf{F}|$ are key steps in both the constrained and original (unconstrained) lift-one algorithms. The theoretical details and functions for the computation of $\mathbf{F}$ are provided in Sections~\ref{sec:model_fisher_glm} and Section~\ref{sec:model_fisher_mlm}.

\subsection{Fisher information for generalized linear models} \label{sec:model_fisher_glm}


The generalized linear model (GLM) broadens the scope of the traditional linear model. In the standard approach, the response variable is expected to change in direct proportion to the covariate. Yet, this isn't always a practical assumption. Take binary outcomes, for instance, the classic linear model falls short here. Similarly, it's unsuitable for positive-only data like count data. \cite{nelder1972generalized} expanded the model to accommodate a wider range of applications.

For a GLM, we assume that response variables $Y_1, \dots, Y_n$ are independent and from the exponential family. Then we have \citep{dobson2018, pmcc1989}
$$E(Y_i \mid \mathbf X_i) = \mu_i,\quad g(\mu_i)=\eta_i=\mathbf X_i^\top \boldsymbol \beta$$ 
where $g$ is a link function, $\boldsymbol \beta = (\beta_1, \beta_2,\dots,\beta_p)^\top$ is the parameter vector of interest, and $\mu_i$ is the conditional expectation of response $Y_i$ given predictors $\mathbf X_i = \mathbf h(\mathbf x_i) = (h_1(\mathbf x_i),\dots, h_p(\mathbf x_i))^\top$, where $i = 1, \dots, n$ with known and deterministic predictor functions $\mathbf h = (h_1, \dots, h_p)^\top$. There are various link functions could be used, for example, logit link $\eta_i = \log\frac{\mu_i}{1-\mu_i}$; probit link $\eta_i = \Phi^{-1}(\mu_i)$, where $\Phi(\cdot)$ is the normal cumulative distribution function; and complementary log-log link $\eta_i = \log\{-\log(1-\mu_i)\}$. Regular linear models can be considered as GLMs with the identity link function and normal responses. Suppose we have a design with $m$ design points $\mathbf x_1, \dots, \mathbf x_m$ that has an exact allocation $(n_1,\dots,n_m)$ where $\sum_i n_i = n$ and corresponding approximate allocation $(w_1,\dots,w_m)=(\frac{n_1}{n},\dots,\frac{n_m}{n})$. The Fisher information matrix $\mathbf F$ under GLMs can be written as \citep{pmcc1989, khuri2006,stufken2012,ymm2016}:
\begin{equation}\label{eq:Fisher_GLM}
\mathbf F = n \mathbf X^\top \mathbf W \mathbf X = n\sum_{i=1}^m w_i \nu_i {\mathbf X}_i{\mathbf X}_i^\top
\end{equation}
where $\mathbf X = (\mathbf X_1, \dots, \mathbf X_m)^\top$ is the $m\times p$ design matrix with $\mathbf X_i = \mathbf h(\mathbf x_i)$, $\mathbf W = \text{diag}\{w_1\nu_1,$ $\dots, w_m\nu_m\}$ is a diagonal matrix with $\nu_i = \frac{1}{\text{Var}(Y_i)} (\frac{\partial \mu_i}{\partial \eta_i})^2$. Here $\nu_i$ represents how much information the design point $\mathbf x_i$ contains. The explicit formats of $\nu_i$ with different response distributions and link functions can be found in Table~5 of the Supplementary Material of \cite{huang2023constrained}. To calculate the Fisher information matrix $\mathbf{F}$ given the approximate allocation $\mathbf{w}$, we may use the \texttt{F\_func\_GLM()} function in the \CRANpkg{CDsampling} package. Additionally, the \texttt{W\_func\_GLM()} function can be used to find the diagonal elements of the matrix $\mathbf{W}$ in the Fisher information matrix \eqref{eq:Fisher_GLM} of GLM. An example of finding the Fisher information matrix for GLM is provided in Example~\ref{ex:GLM_Fisher} of Supplementary Material.

\subsection{Fisher information for multinomial logistic model}\label{sec:model_fisher_mlm}


The multinomial logistic model (MLM) is an extension of GLM aiming to manage responses that fall into multiple categories, such as rating scales and disease severity levels \citep{agresti2013}.

We assume that the responses $\mathbf Y_i = (Y_{i1},\dots,Y_{iJ})$ follow a multinomial distribution with probabilities $(\pi_{i1}, \dots,  \pi_{iJ})$, where $\pi_{ij} = P(Y_i=j \mid {\mathbf x}_i)$, $Y_i \in \{1, \ldots, J\}$ is the $i$th original categorical response, $j = 1,\dots, J$, and $\sum_j \pi_{ij} = 1$. If the response variables are nominal, in other words, there is no natural ordering among different levels, a commonly used MLM is the baseline-category logit model with the $J$th level selected as the baseline category. If the response variable is ordinal, that is,  we have a natural ordering of response levels, there are three commonly used MLMs in the literature:  cumulative logit model, adjacent logit model, and continuation-ratio logit model \citep{bu2020, dousti2023categorical, wang2023identifying}.

In addition to different logit models, the proportional odds (po) assumption is an important concept in MLMs. The po assumption is a parsimonious model assumption, where a model simultaneously uses all $J-1$ logits in a single model with the same coefficients. This means the covariate effect is constant on the odds ratio among different response levels. When the assumption doesn't hold, the model is referred to as a non-proportional odds (npo) model and has more parameters in it. When the assumption only holds for part of the parameters, the model is referred to as a partial proportional odds (ppo) model. 
Commonly used multinomial logit models with po, npo, or ppo assumptions can be summarized in a unified matrix form \citep{pmcc1995, atkinson1999, bu2020}:
$$\mathbf C^\top \log(\mathbf L \boldsymbol \pi_i) = \mathbf X_i \boldsymbol \theta$$
where $$\mathbf C^\top = \begin{pmatrix}
\mathbf I_{J-1} & -\mathbf I_{J-1} & \mathbf 0_{J-1} \\
\mathbf 0^\top_{J-1} & \mathbf 0^\top_{J-1} & 1 \\
\end{pmatrix}$$ 
is a $J \times (2J-1)$ constant matrix, $\mathbf L$ is a $(2J-1)\times J$ constant matrix with different formats among the four different multinomial logit models, and $\boldsymbol \pi_i = (\pi_{i1},\dots,\pi_{iJ})^\top$. The model matrix $\mathbf X_i$ is defined in general as 
$$\mathbf X_i = \begin{pmatrix}
    \mathbf h_1^\top(\mathbf x_i)&  &  & \mathbf h_c^\top(\mathbf x_i)\\
     & \ddots & & \vdots\\
      & & \mathbf h_{J-1}^\top (\mathbf x_i) & \mathbf h_{c}^\top (\mathbf x_i)\\
      \mathbf 0^\top_{p_1} & \hdots & \mathbf 0^\top_{p_{J-1}} & \mathbf 0^\top_{p_c}

\end{pmatrix}_{J \times p}$$
and the parameter $\boldsymbol \theta=(\boldsymbol\beta_1^\top, \dots, \boldsymbol \beta^\top_{J-1}, \boldsymbol\zeta^\top)^\top$ consists of $p=p_1 + \dots + p_{J-1}+p_c$ unknown parameters. Here $\mathbf h_j^\top(\cdot) = (h_{j1}(\cdot),\dots, h_{jp_j}(\cdot))$ are known functions to determine $p_j$ predictors associated with unknown parameters $\boldsymbol \beta_j = (\beta_{j1}, \dots, \beta_{jp_{j}})^\top$ in $j$th response category, and $\mathbf h_c^\top(\cdot) = (h_{1}(\cdot),\dots, h_{p_c}(\cdot))$ are known functions to determine $p_c$ predictors associated with proportional odds parameters $\boldsymbol \zeta = (\zeta_1, \dots, \zeta_{p_c})^\top$. If $\mathbf h_j^\top(\mathbf x_i)\equiv1$, the model is a po model; if $\mathbf h_c^\top(\mathbf x_i)\equiv0$, the model is an npo model. 

According to Theorem 2.1 in \cite{bu2020}, the Fisher information matrix under a multinomial logistic regression model with independent observations can be written as 
\begin{equation}\label{eq:F_sum}
   \mathbf F = \sum_{i=1}^m n_i \mathbf F_i 
\end{equation}
where $\mathbf F_i = (\frac{\partial{\boldsymbol\pi_i}}{\partial{\boldsymbol \theta^\top}})^\top \text{diag}(\boldsymbol \pi_i)^{-1}\frac{\partial \boldsymbol \pi_i}{\partial \boldsymbol \theta^\top}$ with $\frac{\partial \boldsymbol \pi_i}{\partial \boldsymbol \theta^\top} = (\mathbf C^\top \mathbf D_i^{-1} \mathbf L)^{-1}$ and $\mathbf D_i = \text{diag}(\mathbf L \boldsymbol \pi_i)$. Explicit forms of $(\mathbf C^\top \mathbf D_i^{-1} \mathbf L)^{-1}$ can be found in  Section~S.3 of the Supplementary Material of \cite{bu2020}. To calculate the Fisher information matrix $\mathbf F$ for the MLM, one may use the function \texttt{F\_func\_MLM()} in the \CRANpkg{CDsampling} package. An example of finding the Fisher information matrix for MLM is provided in Example~\ref{ex:MLM_Fisher} of the Supplementary Material.

\section{Examples}\label{sec:example}

The methods described in Section~\ref{sec:model} are implemented in the proposed R package \CRANpkg{CDsampling}. The \CRANpkg{CDsampling} package comprises $16$ functions, as detailed in Table~\ref{Tab:summary_functions}. The primary functions of \CRANpkg{CDsampling} package are \texttt{liftone\_constrained\_GLM()} and \texttt{liftone\_constrained\_MLM()}. Additionally, the package includes the original (unconstrained) lift-one algorithm for general experimental designs, accessible via the \texttt{liftone\_GLM()} and \texttt{liftone\_MLM()} functions. Two datasets \texttt{trial\_data} and \texttt{trauma\_data} are provided for illustration purposes.

\begin{table}[htb!]
\scriptsize
\begin{tabularx}{\textwidth}{c|b|m}
\toprule
                        & Usage                                             & Function   \\ \hline
\multirow[c]{8}{*}{\vspace*{-3cm}Model}  & \vspace{0.2cm} Calculating $\mathbf W$ matrix diagonal elements of generalized linear model (see Section~\ref{sec:model_fisher_glm}); providing input for function \texttt{liftone\_GLM()} and \texttt{liftone\_constrained\_GLM()}.\vspace{0.2cm}  & \vspace{0.2cm} \texttt{W\_func\_GLM()} \\
                        &  Calculating Fisher information matrix and its determinant of generalized linear model (see Example~\ref{ex:GLM_Fisher} in Supplementary Material). \vspace{0.2cm} & \texttt{F\_func\_GLM()}, \texttt{Fdet\_func\_GLM()}\\
                        & Calculating Fisher information matrix of multinomial logit model at a specific design point (see Section~\ref{sec:model_fisher_mlm}); using as input of \texttt{liftone\_MLM()} and \texttt{liftone\_constrained\_MLM()}. \vspace{0.2cm} & \texttt{Fi\_func\_MLM()}\\
                        & Calculating Fisher information matrix and its determinant of multinomial logit model (see Example~\ref{ex:MLM_Fisher} in Supplementary Material). \vspace{0.2cm} & \texttt{F\_func\_MLM()}, \texttt{Fdet\_func\_MLM()}\\
                        & Using in \texttt{approxtoexact\_constrained\_func()} to find constrained uniform exact allocation.             \vspace{0.2cm} & \texttt{Fdet\_func\_unif()}\\\hline

\multirow[c]{5}{*}{\vspace*{-3cm} Sampling} & \vspace{0.2cm} Finding unconstrained D-optimal approximate allocation for generalized linear model and multinomial logit model (see Section~\ref{sec:supp_comparison} in Supplementary Material). \vspace{0.2cm}& \vspace{0.2cm} \texttt{liftone\_GLM()}, \texttt{liftone\_MLM()}\\
                        & Finding constrained D-optimal approximate allocation for generalized linear model and multinomial logit model (see Section~\ref{sec:example}). \vspace{0.2cm} & \texttt{liftone\_constrained\_GLM()}, \texttt{liftone\_constrained\_MLM()}\\
                        & Transferring approximate allocation to exact allocation (see Section~\ref{sec:example}). \vspace{0.2cm} & \texttt{approxtoexact\_constrained\_func()}\\
                        &                                                   & \texttt{approxtoexact\_func()}\\
                        & Finding constrained uniform exact allocation for bounded constraint (see Section~\ref{sec:example_glm}). \vspace{0.2cm} & \texttt{bounded\_uniform()}\\\hline
\multirow[c]{1}{*}{\vspace*{-1cm} Example Specific} & \vspace{0.2cm} Finding $I$ set for exact allocation conversion in \texttt{trial\_data} and \texttt{trauma\_data} examples (see Section~\ref{sec:example} and Section~\ref{sec:supp_iset_codes} in Supplementary Material). \vspace{0.2cm} & \vspace{0.2cm} \texttt{iset\_func\_trauma()}, \texttt{iset\_func\_trial()}\\\bottomrule
\end{tabularx}
\caption{Functions and corresponding usages in the \CRANpkg{CDsampling} package.}\label{Tab:summary_functions}
\end{table}

In the remainder of this section, we present two examples to illustrate the usage of the \CRANpkg{CDsampling} package for sampling problems in paid research studies, using datasets provided by \CRANpkg{CDsampling}. All results in this section were generated using R version 4.3.2 on a macOS Sonoma 14.6.1 system.

\subsection{Applications in paid research study: $\tt trial\_data$ example}\label{sec:example_glm}

\begin{figure}[htb!]
    \centering
    \includegraphics[width=0.6\textwidth]{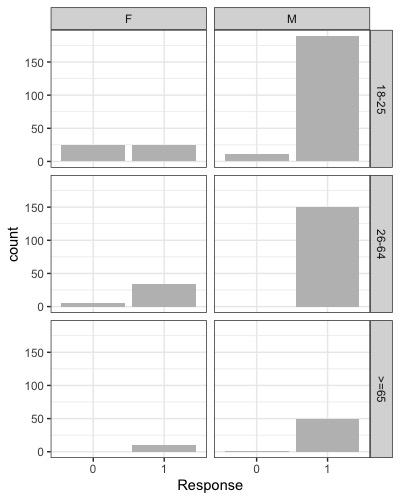}
    \caption{The number of patients from different gender (F, M) and age groups ($18-25$, $26-64$, and $\ge 65$) and their responses  ($0$ indicating ineffectiveness, $1$ indicating effectiveness) to treatment in \texttt{trial\_data} of \CRANpkg{CDsampling} package.}
    \label{fig:trial_data}
\end{figure}

The \texttt{trial\_data} dataset is a simulated data for a toy example of paid research studies. This study includes a cohort of $500$ patients for a clinical trial, with gender and age as stratification factors. A logistic regression model incorporates these factors as covariates: {\tt gender} (coded as $0$ for female and $1$ for male) and {\tt age} (coded as two dummy variables $age\_1$ and $age\_2$ with $(age\_1, age\_2) = (0, 0)$ for age group $18\sim25$; $(1, 0)$ for age group $26\sim64$, and $(0, 1)$ for age group $65$ and above). For simplicity, the study assumes binary gender options and a tripartite age categorization. In total, there are $m=6$ combinations of covariate factors. In practice, non-binary gender options and a ``prefer not to answer'' choice may be included to respect gender diversity and protect patient confidentiality. The response $Y$ denotes the treatment's efficacy ($0$ indicating ineffectiveness, $1$ indicating effectiveness). The data is generated by the logistic regression model 
\begin{equation}\label{eq:trial_logistic_model}
{\rm logit} \{P(Y_{ij}=1 \mid x_{gender\_i}, x_{age\_1i}, x_{age\_2i})\} = \beta_0 + \beta_1 x_{gender\_i} + \beta_{21} x_{age\_1i} + \beta_{22} x_{age\_2i} 
\end{equation}
with $(\beta_0, \beta_1, \beta_{21}, \beta_{22}) = (0,3,3,3)$, where $i=1, \ldots, 6$ stands for the $i$th covariate combination, and $j = 1, \ldots, N_i$ is an index of patients who fall into the $i$th covariate combination or sampling subgroups. Figure~\ref{fig:trial_data} illustrates the distribution of treatment efficacy across different {\tt gender} and {\tt age} groups.

In this example, it is posited that a sample of $n=200$ participants is desired from the population of $N=500$ volunteers due to budget constraints. The objective is to examine the variation in efficacy rates across gender and age demographics. Should a pilot study or relevant literature provide approximate values for the model parameters, a constrained lift-one algorithm may be employed to find a locally D-optimal design. Conversely, if only partial parameter information is available, the expectation can be deduced from some prior distributions, and the constrained lift-one algorithm can be utilized to determine an EW D-optimal allocation (substituting $\mathbf W$ in the Fisher information matrix \eqref{eq:Fisher_GLM} with $E(\mathbf W)$). 

There are $m=6$ design points, corresponding to gender and age group combinations $(x_{gender\_i}, $ $x_{age\_1i}, x_{age\_2i}) = (0,0,0)$, $(0,1,0)$, $(0,0,1)$, $(1,0,0)$, $(1,1,0)$, and $(1,0,1)$, respectively. The numbers of available volunteers in the six categories are $(N_1, N_2, \ldots, N_6) = (50, 40, 10, $ $200, 150, 50)$. Our goal is to find the constrained D-optimal allocation $(w_1, w_2, \dots, w_6)$ in the set of all feasible allocations $S = \{(w_1, \ldots,$ $ w_m)^T \in S_0 \mid n w_i \leq N_i, i=1, \ldots, m\}$. 

We consider the logistic regression model \eqref{eq:trial_logistic_model}, to find the locally D-optimal design, we assume, for illustrative purposes, that the model parameters $(\beta_0, \beta_{1}, \beta_{21}, \beta_{22})$ $=(0,3,3,3)$. We may define the parameters and the model matrix as follows. Subsequently, we find the $\mathbf W$ matrix in \eqref{eq:Fisher_GLM} for the Fisher information matrix $\mathbf F$. 

\begin{example}
> beta = c(0, 3, 3, 3) #coefficients
> #design matrix
> X=matrix(data=c(1,0,0,0,1,0,1,0,1,0,0,1,1,1,0,0,1,1,1,0,1,1,0,1), ncol=4, byrow=TRUE) 
> W=W_func_GLM(X=X, b=beta, link="logit") #find W as input of constrained liftone
\end{example}
To define the number of design points, sample size, and constraints with $S$, we use the following R codes (see Section~S3 in the Supplementary Material of \cite{huang2023constrained} for details on finding $r_{i1}$ and $r_{i2}$ in step~3 of the constrained liftone algorithm in Figure~\ref{fig:constrained_liftone_algo}): 

\begin{example}
> rc = c(50, 40, 10, 200, 150, 50)/200 #available volunteers/sample size
> m = 6 #design points
    
> g.con = matrix(0,nrow=(2*m+1), ncol=m) #constraints
> g.con[1,] = rep(1, m)
> g.con[2:(m+1),] = diag(m)
> g.con[(m+2):(2*m+1), ] = diag(m)
> g.dir = c("==", rep("<=", m), rep(">=", m)) #direction
> g.rhs = c(1, rc, rep(0, m)) #righ-hand side

> #lower bound in step 3 of constrained liftone    
> lower.bound=function(i, w){ 
+   rc = c(50, 40, 10, 200, 150, 50)/200
+   m=length(w) 
+   temp = rep(0,m)
+   temp[w>0]=1-pmin(1,rc[w>0])*(1-w[i])/w[w>0];
+   temp[i]=0;
+   max(0,temp);
+   } 

> #upper bound in step 3 of constrained liftone
> upper.bound=function(i, w){ 
+   rc = c(50, 40, 10, 200, 150, 50)/200
+   min(1,rc[i]);
+   } 
\end{example}

To identify the subgroups of the output D-optimal allocations, we may add an optional label for each of the $m=6$ covariatres combination or subgroups as ``F, 18-25'',  ``F, 26-64'', ``F, >=65'', ``M, 18-2'', ``M, 26-6'', ``M, >=65'' using the following codes: 

\begin{example}
> label = c("F, 18-25", "F, 26-64", "F, >=65", "M, 18-25", "M, 26-64", "M, >=65")
\end{example}

Then, we run the constrained lift-one algorithm with \texttt{\seqsplit{liftone\_constrained\_GLM()}} to find the constrained D-optimal approximate allocation. 
\begin{example}
> set.seed(092)
> approximate_design = liftone_constrained_GLM(X=X, W=W, g.con=g.con, g.dir=g.dir, 
+ g.rhs=g.rhs, lower.bound=lower.bound, upper.bound=upper.bound, reltol=1e-10, 
+ maxit=100, random=TRUE, nram=4, w00=NULL, epsilon=1e-8, label=label)
\end{example}
The design output is presented below: 

\begin{example}
> print(approximate_design)

Optimal Sampling Results:
================================================================================
Optimal approximate allocation:
   F, 18-25 F, 26-64 F, >=65 M, 18-25 M, 26-64 M, >=65
w  0.25     0.2      0.05    0.5      0.0      0.0    
w0 0.25     0.0      0.05    0.7      0.0      0.0    
--------------------------------------------------------------------------------
maximum :
2.8813e-08 
--------------------------------------------------------------------------------
convergence :
TRUE 
--------------------------------------------------------------------------------
itmax :
9.0 
--------------------------------------------------------------------------------
deriv.ans :
0.0, 3.6017e-08, 4.8528e-07, -1.1525e-07, -1.0310e-07, -7.9507e-08 
--------------------------------------------------------------------------------
gmax :
0.0 
--------------------------------------------------------------------------------
reason :
"gmax <= 0" 
--------------------------------------------------------------------------------
\end{example}

\noindent The output includes several key components: 
\begin{itemize}
    \item $\mathbf w$: reports the D-optimal approximate allocation.
    \item $\mathbf w_0$: reports the random initial approximate allocation used to initialize optimization.
    \item \textbf{maximum}: reports the maximum determinant of Fisher information matrix.
    \item \textbf{reason}: reports the termination criterion for the constrained lift-one algorithm including either ``all derivative $\le$ 0'' or ``gmax $\le$ 0'', which corresponds to step~$7^\circ$ and step~$8^\circ$ in the constrained lift-one algorithm in Figure~\ref{fig:constrained_liftone_algo}. 
\end{itemize}

In practical terms, exact allocations are more beneficial. One may use the constrained approximate to exact allocation algorithm depicted in Figure~\ref{fig:approx_to_exact_algo}, which is implemented as the \texttt{\seqsplit{approxtoexact\_constrained\_func()}} function.

\begin{example}
> exact_design = approxtoexact_constrained_func(n=200, w=approximate_design$w, m=6, 
+ beta=beta, link='logit', X=X, Fdet_func=Fdet_func_GLM, iset_func=iset_func_trial,
+ label=label)    
> print(exact_design)

Optimal Sampling Results:
================================================================================
Optimal exact allocation:
                F, 18-25 F, 26-64 F, >=65 M, 18-25 M, 26-64 M, >=65
allocation      50.0     40.0     10.0    100.0    0.0      0.0    
allocation.real 0.25     0.2      0.05    0.5      0.0      0.0    
--------------------------------------------------------------------------------
det.maximum :
46.1012 
--------------------------------------------------------------------------------
\end{example}

\noindent The output provides three key components for the sampling results: 
\begin{itemize}
    \item \textbf{allocation}: reports the exact allocation of D-optimal sampling, specifying the number of subjects to sample from each subgroup. 
    \item \textbf{allocation.real}: reports the real-number approximate allocation used prior to integer conversion. 
    \item \textbf{det.maximum}: reports the maximum determinant of the Fisher information matrix by the optimal design. 
\end{itemize}

In this example, the D-optimal exact allocation is to sample $50$ subjects from the ``female, $18-25$'' subgroup, $40$ subjects from the ``female, $26-64$'' subgroup, $10$ subjects from the ``female, $\ge 65$'' subgroup, and $100$ subjects from the ``male, $18-25$'' subgroup. Such a design may not explore all the design space, and lead to extreme design cases. In practice, allocating some subjects to the omitted subgroups ``male, $26-64$'' and ``male, $\ge 65$'' subgroups could improve the robustness and reduce the risk of overfitting. 


Alternatively, one may aim for EW D-optimal allocations when partial coefficients information available with the ${\mathbf W}$ matrix substituted by the expectation (EW stands for expected weighted). To calculate these expectations, one may define prior distributions for the parameters based on available information. For instance, in this scenario, we assume the following independent prior distributions: $\beta_0\sim$ uniform$(-2,2)$, $\beta_1\sim$ uniform$(-1,5)$, $\beta_{21}\sim$ uniform$(-1,5)$, and $\beta_{22}\sim$ uniform(-1, 5). Subsequently, the diagonal elements of ${\mathbf W}$ are determined through integration. For $i=1,\dots,m$ and $\eta_i = \beta_0 + x_{gender\_i}\beta_1+ x_{age\_i1}\beta_{21}+ x_{age\_i2}\beta_{22}$, we calculate the key component $E(\nu_i)$ of the $i$th diagonal element of ${\mathbf W}$ through: 
\begin{equation*}
  \int_{-1}^{5} \int_{-1}^{5} \int_{-1}^{5} \int_{-2}^{2} \frac{\exp(\eta_i)}{(1+\exp(\eta_i))^2} {\rm Pr}(\beta_0){\rm Pr}(\beta_{1}){\rm Pr}(\beta_{21}){\rm Pr}(\beta_{22}) \, d\beta_0 d\beta_1 d\beta_{21} d\beta_{22} 
\end{equation*}
where $\rm{Pr}(\cdot)$ stands for the corresponding probability density function.
We use the \texttt{hcubature()} function in the \CRANpkg{cubature} package to calculate the integration as illustrated by the R codes below.   

\begin{example}
> unif.prior <- rbind(c(-2, -1, -1, -1), c(2,  5,  5, 5)) #prior parameters

> #find expectation of W matrix given priors
> W.EW.unif = matrix(rep(0,6)) 
> for (i in 1:6){
+  x = X[i,]
+  W.EW.unif[i] = hcubature(function(beta) dunif(beta[1], min=unif.prior[1,1], 
max=unif.prior[2,1])*dunif(beta[2], min=unif.prior[1,2], max=unif.prior[2,2])*
dunif(beta[3], min=unif.prior[1,3], max=unif.prior[2,3])*dunif(beta[4], 
min=unif.prior[1,4], max=unif.prior[2,4])*(exp(x[1]*beta[1]+x[2]*beta[2]+x[3]*
beta[3]+x[4]*beta[4])/(1+exp(x[1]*beta[1]+x[2]*beta[2]+x[3]*beta[3]+x[4]*beta[4]))^2), 
lowerLimit = unif.prior[1,], upperLimit  = unif.prior[2,])$integral
+  }
\end{example}
Given the expectation of ${\mathbf W}$, the functions \texttt{liftone\_constrained\_GLM()} and \texttt{\seqsplit{approxtoexact\_constrained\_func()}} are used for deriving the constrained EW D-optimal approximate allocation and the corresponding exact allocation, respectively. This process follows a similar procedure to that used for local D-optimal approximate allocation. 

\begin{example}
> set.seed(123)
> approximate_design_EW = liftone_constrained_GLM(X=X, W=W.EW.unif,  g.con=g.con, 
+ g.dir=g.dir, g.rhs=g.rhs, lower.bound=lower.bound, upper.bound=upper.bound, 
+ reltol=1e-12, maxit=100, random=TRUE, nram=4, w00=NULL, epsilon=1e-10, label=label)

> exact_design = approxtoexact_constrained_func(n=200, 
+ w=approximate_design_EW$w, m=6, beta=beta, link='logit', X=X, Fdet_func=Fdet_func_GLM, 
+ iset_func=iset_func_trial, label=label)
\end{example}
The output is summarized with \texttt{print()} function and presented below: 
\begin{example}
> print(exact_design_EW)

Optimal Sampling Results:
================================================================================
Optimal exact allocation:
                F, 18-25 F, 26-64 F, >=65 M, 18-25 M, 26-64 M, >=65
allocation      48.0     40.0     10.0    43.0     19.0     40.0   
allocation.real 0.2406   0.2      0.05    0.2102   0.0991   0.2001 
--------------------------------------------------------------------------------
det.maximum :
25.59 
--------------------------------------------------------------------------------
\end{example}


In situations when the model parameters are unknown, the constrained uniform allocation is applicable. This method entails sampling an equal number of patients from each category within the given constraints. The selection criterion is $n_i = \min\{k, N_i\}$ or $\min\{k, N_i\}+1$ with $k$ satisfying $\sum_{i=1}^m \min\{k, N_i\} \leq n < \sum_{i=1}^m \min\{k+1, N_i\}$, where $N_i$ represents the maximum allowable number for each category. This is an example of a bounded design problem, where each category has an upper boundary. The function \texttt{bounded\_uniform()} can be used to find the constrained uniform allocation with the \texttt{trial\_data} example and ``\textbf{allocation}'' in the output representing the constrained uniform allocation.  

\begin{example}
> bounded_uniform(Ni=c(50, 40, 10, 200, 150, 50), nsample=200, label=label)

Optimal Sampling Results:
================================================================================
Optimal exact allocation:
           F, 18-25 F, 26-64 F, >=65 M, 18-25 M, 26-64 M, >=65
allocation 38.0     38.0     10.0    38.0     38.0     38.0   
--------------------------------------------------------------------------------
\end{example}

Alternatively, we may also use \texttt{approxtoexact\_constrained\_func()} to find the same constrained uniform exact allocation. This function can be used under fairly general constraints. To find the constrained uniform exact allocation using \texttt{\seqsplit{approxtoexact\_constrained\_func()}}, we suggest starting with one subject in each stratum or subgroup, which corresponds to the approximate allocation $\mathbf{w_{00}}=(1/200,1/200,1/200,1/200,1/200,1/200)$ in this case. 

\begin{example}
> w00 = rep(1/200, 6) #initial approximate allocation
> unif_design = approxtoexact_constrained_func(n=200, w=w00, m=6, beta=NULL, 
+ link=NULL, X=NULL, Fdet_func=Fdet_func_unif, iset_func=iset_func_trial, label=label)

> print(unif_design)

Optimal Sampling Results:
================================================================================
Optimal exact allocation:
                F, 18-25 F, 26-64 F, >=65 M, 18-25 M, 26-64 M, >=65
allocation      38.0     38.0     10.0    38.0     38.0     38.0   
allocation.real 0.005    0.005    0.005   0.005    0.005    0.005  
--------------------------------------------------------------------------------
det.maximum :
792351680.0 
--------------------------------------------------------------------------------

\end{example}

The \texttt{iset\_func\_trial()} function in the \CRANpkg{CDsampling} package is specifically designed for the \texttt{trial\_data} example, which defines the set $I$ in step~$1^\circ$ and step~$2.4$ in the constrained approximate to exact algorithm depicted by Figure~\ref{fig:approx_to_exact_algo}. This function serves as a template that users can adapt to their specific constraints by modifying the codes. The package includes two such template functions: \texttt{\seqsplit{\textbf{iset\_func\_trial()}}} and \texttt{\seqsplit{\textbf{iset\_func\_trauma()}}} with details provided in Section~\ref{sec:supp_iset_codes} of Supplementary Material. 

To perform a comparison analysis on different sampling strategies, including the constrained D-optimal allocation, the constrained EW D-optimal allocation with uniform priors, the constrained uniform allocation, simple random sample without replacement (SRSWOR), as well as the full data (all the $500$ patients enrolled), we simulate their responses $Y_{ij}$'s based on model~\eqref{eq:trial_logistic_model} with parameter values $(0,3,3,3)$. We apply each sampling strategy to obtain a sample of $n=200$ observations out of $500$, and estimate the parameters using the $200$ observations. The exception is the full data method, where estimation is performed using all $500$ patients. We use the root mean square error (RMSE) to measure the accuracy of the estimates (see Section~4 of \cite{huang2023constrained} for more theoretical and technical details). We repeat the procedure $100$ times and display the corresponding RMSEs in Figure~\ref{fig:RMSE_GLM} and simulation codes in Supplementary Material Section~\ref{sec:supp_simulation} \citep{ggplot2, dplyr}. Obviously, if we use the full data (all $500$ patients) to fit the model, its RMSE attains the lowest. Besides that, the constrained locally D-optimal allocation and the constrained EW D-optimal allocation have a little higher RMSEs than the full data estimates, but outperform SRSWOR and the constrained uniform allocation. The sampling strategy based on the constrained uniform allocation doesn't need any model information and is a more robust sampling scheme, which is still better than SRSWOR. 

\begin{figure}[htb!]
    \centering
    \includegraphics[width=0.95\textwidth]{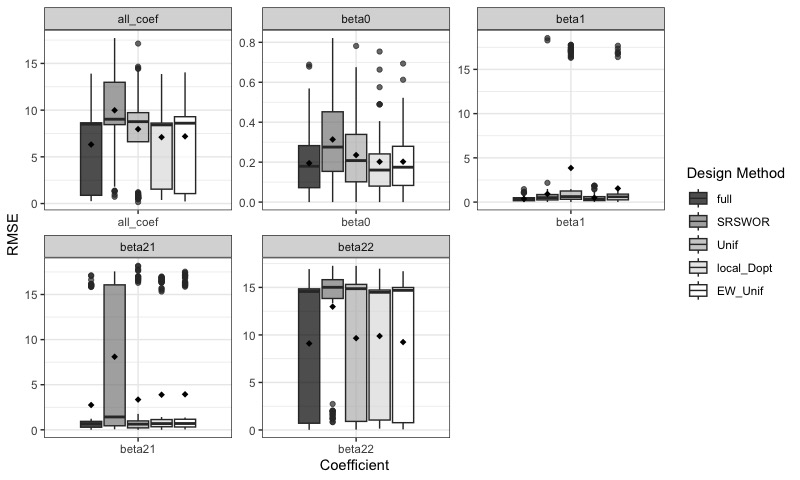}
    \caption{Boxplots of RMSEs obtained from $100$ simulations  using full data (full), SRSWOR, constrained uniform design (Unif), the constrained locally D-optimal allocation (local\_Dopt), and constrained EW D-optimal allocation with uniform priors (EW\_Unif), with black diamonds representing average RMSE.}
    \label{fig:RMSE_GLM}
\end{figure}

\subsection{Applications in paid research study: $\tt trauma\_data$ example}\label{sec:MLM_example}
\begin{figure}[htb!]
    \centering
    \includegraphics[width=0.95\textwidth]{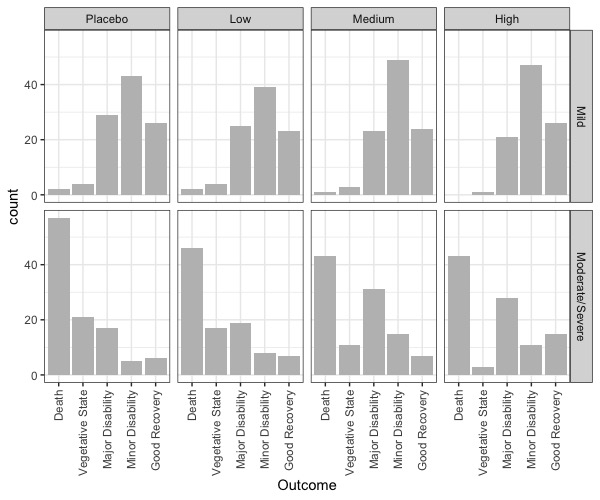}
    \caption{The number of patients from different {\tt dose} levels (Placebo, Low, Medium, and High) and {\tt severity} grades (Mild, Moderate/Severe) groups and their treatment outcomes in \texttt{trauma\_data} of \CRANpkg{CDsampling} package.}
    \label{fig:trauma_data}
\end{figure}

In the \CRANpkg{CDsampling} package, \texttt{trauma\_data} is a dataset of $N=802$ trauma patients from \cite{chuang1997}, stratified according to the trauma severity at the entry time of the study with $392$ mild and $410$ moderate/severe patients enrolled. The study involved four treatment groups determined by the {\tt dose} level, $x_{i1} = 1$ ({\tt Placebo}), $2$ ({\tt Low dose}), $3$ ({\tt Medium dose}), and $4$ ({\tt High dose}). Combining with {\tt severity} grade ($x_{i2} = 0$ for {\tt mild} or $1$ for {\tt moderate}/{\tt severe}), there are $m=8$ distinct experimental settings with $(x_{i1}, x_{i2}) = (1,0), (2,0), (3,0), (4,0), (1,1), (2,1), (3,1), (4,1)$, respectively. The responses belong to five ordered categories, {\tt Death} ($1$), {\tt Vegetative state} ($2$), {\tt Major disability} ($3$), {\tt Minor disability} ($4$) and {\tt Good} {\tt recovery} ($5$), known as the Glasgow Outcome Scale \citep{jennett1975}. Figure~\ref{fig:trauma_data} shows the distribution of outcomes over {\tt severity} grades and {\tt dose} levels. 
In this example, we have $m=8$ subgroups, which are combinations of the two covariates categories: \texttt{dose} levels and \texttt{severity} grades. We aim to enroll $n=600$ patients from the $802$ available patients. The collection of feasible allocations is inherently constrained by the number of patients in different severity grades, defined as $S=\{(w_1, \ldots, w_8)^\top \in S_0 \mid n(w_1+w_2+w_3+w_4) \leq 392, n(w_5+w_6+w_7+w_8) \leq 410\}$. The constraints specify that in the sample, the number of patients with mild symptoms must not exceed $392$ across all dose levels, while those with moderate/severe symptoms must not exceed $410$.

The parameters fitted from the \texttt{trauma\_data} are $\boldsymbol\beta = (\hat\beta_{11}, \hat\beta_{12}, \hat\beta_{13}, \hat\beta_{21}, \hat\beta_{22}, \hat\beta_{23}, \hat\beta_{31}, \hat\beta_{32}, \hat\beta_{33}, \hat\beta_{41},$ $\hat\beta_{42}, \hat\beta_{43})^\top  = (-4.047, -0.131, 4.214, -2.225, -0.376, 3.519, -0.302, -0.237,  2.420, 1.386,  -0.120,  1.284)^\top$. The model can be written in the following format:
    \begin{eqnarray*}
    \log\left(\frac{\pi_{i1}}{\pi_{i2}+\dots+\pi_{i5}}\right) &=& \beta_{11}+\beta_{12}x_{i1}+\beta_{13}x_{i2}\\
    \log\left(\frac{\pi_{i1}+\pi_{i2}}{\pi_{i3}+\pi_{i4}+\pi_{i5}}\right) &=& \beta_{21}+\beta_{22}x_{i1}+\beta_{23}x_{i2}\\    \log\left(\frac{\pi_{i1}+\pi_{i2}+\pi_{i3}}{\pi_{i4}+\pi_{i5}}\right) &=& \beta_{31}+\beta_{32}x_{i1}+\beta_{33}x_{i2}\\
    \log\left(\frac{\pi_{i1}+\dots+\pi_{i4}}{\pi_{i5}}\right) &=&\beta_{41}+\beta_{42}x_{i1}+\beta_{43}x_{i2}
    \end{eqnarray*}
    where $i=1,\dots,8$.

We use the R codes below to define the model matrix and coefficients.

\begin{example}
> J=5; p=12; m=8; #response levels; parameters; subgroups

> #coefficients
> beta = c(-4.047, -0.131, 4.214, -2.225, -0.376, 3.519, -0.302, -0.237,  2.420, 1.386,  
+ -0.120,  1.284)

> #define design matrix of 8 subgroups
> Xi=rep(0,J*p*m); dim(Xi)=c(J,p,m);
> Xi[,,1] = rbind(c( 1, 1, 0, 0, 0, 0, 0, 0, 0, 0, 0, 0), 
+           c( 0, 0, 0, 1, 1, 0, 0, 0, 0, 0, 0, 0), c( 0, 0, 0, 0, 0, 0, 1, 1, 0, 0, 0, 0), 
+           c( 0, 0, 0, 0, 0, 0, 0, 0, 0, 1, 1, 0), c( 0, 0, 0, 0, 0, 0, 0, 0, 0, 0, 0, 0))

> Xi[,,2] = rbind(c( 1, 2, 0, 0, 0, 0, 0, 0, 0, 0, 0, 0), 
+           c( 0, 0, 0, 1, 2, 0, 0, 0, 0, 0, 0, 0), c( 0, 0, 0, 0, 0, 0, 1, 2, 0, 0, 0, 0), 
+           c( 0, 0, 0, 0, 0, 0, 0, 0, 0, 1, 2, 0), c( 0, 0, 0, 0, 0, 0, 0, 0, 0, 0, 0, 0))

> Xi[,,3] = rbind(c( 1, 3, 0, 0, 0, 0, 0, 0, 0, 0, 0, 0), 
+           c( 0, 0, 0, 1, 3, 0, 0, 0, 0, 0, 0, 0), c( 0, 0, 0, 0, 0, 0, 1, 3, 0, 0, 0, 0), 
+           c( 0, 0, 0, 0, 0, 0, 0, 0, 0, 1, 3, 0), c( 0, 0, 0, 0, 0, 0, 0, 0, 0, 0, 0, 0))

> Xi[,,4] = rbind(c( 1, 4, 0, 0, 0, 0, 0, 0, 0, 0, 0, 0),
+           c( 0, 0, 0, 1, 4, 0, 0, 0, 0, 0, 0, 0), c( 0, 0, 0, 0, 0, 0, 1, 4, 0, 0, 0, 0), 
+           c( 0, 0, 0, 0, 0, 0, 0, 0, 0, 1, 4, 0), c( 0, 0, 0, 0, 0, 0, 0, 0, 0, 0, 0, 0))

> Xi[,,5] = rbind(c( 1, 1, 1, 0, 0, 0, 0, 0, 0, 0, 0, 0), 
+           c( 0, 0, 0, 1, 1, 1, 0, 0, 0, 0, 0, 0), c( 0, 0, 0, 0, 0, 0, 1, 1, 1, 0, 0, 0), 
+           c( 0, 0, 0, 0, 0, 0, 0, 0, 0, 1, 1, 1), c( 0, 0, 0, 0, 0, 0, 0, 0, 0, 0, 0, 0))

> Xi[,,6] = rbind(c( 1, 2, 1, 0, 0, 0, 0, 0, 0, 0, 0, 0), 
+           c( 0, 0, 0, 1, 2, 1, 0, 0, 0, 0, 0, 0), c( 0, 0, 0, 0, 0, 0, 1, 2, 1, 0, 0, 0), 
+           c( 0, 0, 0, 0, 0, 0, 0, 0, 0, 1, 2, 1), c( 0, 0, 0, 0, 0, 0, 0, 0, 0, 0, 0, 0))

> Xi[,,7] = rbind(c( 1, 3, 1, 0, 0, 0, 0, 0, 0, 0, 0, 0), 
+           c( 0, 0, 0, 1, 3, 1, 0, 0, 0, 0, 0, 0), c( 0, 0, 0, 0, 0, 0, 1, 3, 1, 0, 0, 0), 
+           c( 0, 0, 0, 0, 0, 0, 0, 0, 0, 1, 3, 1), c( 0, 0, 0, 0, 0, 0, 0, 0, 0, 0, 0, 0))

> Xi[,,8] = rbind(c( 1, 4, 1, 0, 0, 0, 0, 0, 0, 0, 0, 0), 
+           c( 0, 0, 0, 1, 4, 1, 0, 0, 0, 0, 0, 0), c( 0, 0, 0, 0, 0, 0, 1, 4, 1, 0, 0, 0), 
+           c( 0, 0, 0, 0, 0, 0, 0, 0, 0, 1, 4, 1), c( 0, 0, 0, 0, 0, 0, 0, 0, 0, 0, 0, 0))
\end{example}

To define the sample size, the constraints, and the functions of lower and upper boundaries $r_{i1}$ and $r_{i2}$, we may use the following R codes (see Section~S3 in the Supplementary Material of \cite{huang2023constrained} for details on finding $r_{i1}$ and $r_{i2}$): 
\begin{example}
> nsample=600 #sample size
> constraint = c(392, 410)  #mild:severe

> #lower bound function in step 3 of constrained liftone
> lower.bound <- function(i, w0){
+   n = 600
+   constraint = c(392,410)
+   if(i <= 4){
+     a.lower <- (sum(w0[5:8])-(constraint[2]/n)*(1-w0[i]))/(sum(w0[5:8]))}
+   else{
+     a.lower <- (sum(w0[1:4])-(constraint[1]/n)*(1-w0[i]))/(sum(w0[1:4]))}
+    a.lower}

> #upper bound function in step 3 of constrained liftone
> upper.bound <- function(i, w0){
+  n = 600
+  constraint = c(392,410)
+  if(i <= 4){
+    b.upper <- ((constraint[1]/n)*(1-w0[i]) - (sum(w0[1:4])-w0[i]))/(1-sum(w0[1:4]))}
+  else{
+    b.upper <- ((constraint[2]/n)*(1-w0[i]) - (sum(w0[5:8])-w0[i]))/(1-sum(w0[5:8]))}
+    b.upper}

> #define constraints
> g.con = matrix(0,nrow=length(constraint)+1+m, ncol=m)
> g.con[2:3,] = matrix(data=c(1,1,1,1,0,0,0,0,0,0,0,0,1,1,1,1), ncol = m, byrow=TRUE)
> g.con[1,] = rep(1, m)
> g.con[4:(length(constraint)+1+m), ] = diag(1, nrow=m)
> g.dir = c("==", "<=","<=", rep(">=",m))
> g.rhs = c(1, ifelse((constraint/nsample<1),constraint/nsample,1), rep(0, m))
\end{example}

Then, we may define an optional label of the sampling subgroups that corresponds to each of the $m=8$ subgroups using the following code: 

\begin{example}
> label=label = c("Placebo-Mild", "Low-Mild", "Medium-Mild", "High-Mild", "Placebo-Severe", 
+ "Low-Severe", "Medium-Severe", "High-Severe")
\end{example}


We then run the constrained lift-one algorithm to find the constrained D-optimal approximate allocation using \texttt{liftone\_constrained\_MLM()} function and convert the approximate allocation to an exact allocation with \texttt{approxtoexact\_constrained\_func} function.
\begin{example}
> set.seed(123)
> approx_design = liftone_constrained_MLM(m=m, p=p, Xi=Xi, J=J, beta=beta, 
+ lower.bound=lower.bound, upper.bound=upper.bound, g.con=g.con, g.dir=g.dir, 
+ g.rhs=g.rhs, w00=NULL, link='cumulative',  Fi.func=Fi_func_MLM, reltol=1e-5, 
+ maxit=500, delta=1e-6, epsilon=1e-8, random=TRUE, nram=3, label=label)

> exact_design = approxtoexact_constrained_func(n=600, w=approx_design$w, m=8, 
+ beta=beta, link='cumulative', X=Xi, Fdet_func=Fdet_func_MLM, 
+ iset_func=iset_func_trauma, label=label)

> print(exact_design)

Optimal Sampling Results:
================================================================================
Optimal exact allocation:
             Placebo-Mild Low-Mild Medium-Mild High-Mild Placebo-Severe
allocation      155.0        0.0      0.0         100.0     168.0
allocation.real 0.2593       0.0      0.0         0.1667    0.2796
              Low-Severe Medium-Severe High-Severe
allocation      0.0        0.0           177.0
allocation.real 0.0        0.0           0.2944
--------------------------------------------------------------------------------
det.maximum :
1.63163827059162e+23
--------------------------------------------------------------------------------
\end{example}

The \textbf{allocation} output provides the exact allocation of the sampling across different treatment-severity subgroups, representing the implementable sample sizes for each subgroup. The result is derived by converting the \textbf{allocation.real}, which is the D-optimal approximate allocation outcome from \texttt{liftone\_constrained\_MLM()}. 

 As the \texttt{trauma\_data} example doesn't have bounded constraints,  to find the constrained uniform sampling allocation, we use \texttt{approxtoexact\_constrained\_func()} with one subject in each stratum or subgroup as the input, that is, the approximate allocation $\mathbf w = (1/600, 1/600, 1/600, 1/600,$ $1/600, 1/600, 1/600, 1/600)$. The corresponding $I$ set function is provided in the \CRANpkg{CDsampling} package, and it can be easily defined according to other constraints, see Section~\ref{sec:supp_iset_codes} of Supplementary Material. Note that the determinant provided by \texttt{approxtoexact\_constrained\_func()} for different designs are not comparable, as the criteria \texttt{Fdet\_func} differ.

\begin{example}
> unif_design = approxtoexact_constrained_func(n=600, w=rep(1/600,8), m=8, 
+ beta=NULL, link=NULL, X=NULL, Fdet_func=Fdet_func_unif, iset_func=iset_func_trauma)

> print(unif_design)

Optimal Sampling Results:
================================================================================
Optimal exact allocation:
                Placebo-Mild Low-Mild Medium-Mild High-Mild Placebo-Severe Low-Severe
allocation      75.0         75.0     75.0        75.0      75.0           75.0      
allocation.real 0.0017       0.0017   0.0017      0.0017    0.0017         0.0017    
                Medium-Severe High-Severe
allocation      75.0          75.0       
allocation.real 0.0017        0.0017     
--------------------------------------------------------------------------------
det.maximum :
1001129150390625 
--------------------------------------------------------------------------------
\end{example}

\section{Summary}

The current version of \CRANpkg{CDsampling} implements D-optimal allocations within both paid research sampling and general study frameworks with or without constraints. Its primary objective is to optimize sampling allocations for better model estimation accuracy in the studies. The package includes \texttt{F\_func\_GLM()} and \texttt{F\_func\_MLM()} for the computation of the Fisher information matrix of GLMs and MLMs, respectively. It is noteworthy that standard linear regression models are special GLM with an identity link function and Gaussian-distributed responses, which is also supported by our package. Theoretical results are summarized in Section~\ref{sec:model_fisher_glm} and Section~\ref{sec:model_fisher_mlm} while illustrative examples are provided in Supplementary Section~\ref{sec:supp_fisher_example}. 

To find standard or unconstrained D-optimal allocations, our package implements the lift-one algorithm through functions \texttt{liftone\_GLM()} and \texttt{liftone\_MLM()}. Paid research studies often impose sampling constraints. To address this, the constrained lift-one algorithm can be applied using functions \texttt{liftone\_constrained\_GLM()} and \texttt{liftone\_constrained\_MLM()}. An example illustrating the difference between the lift-one algorithm and the constrained lift-one algorithm is provided in Supplementary Section~\ref{sec:supp_comparison} while Section~\ref{sec:example} presents two application examples from paid research studies. 

In the absence of model information, \texttt{constrained\_uniform()} function is available to find a robust constrained uniform allocation with bounded constraints, while the \texttt{\seqsplit{approxtoexact\_constrained\_func()}} function can be used to find constrained uniform allocation with more general constraints. For transitioning from approximate to exact allocations, the package provides \texttt{approxtoexact\_constrained\_func()} for constrained cases and \texttt{\seqsplit{approxtoexact\_func()}} for unconstrained cases. Detailed applications for both GLMs and MLMs are provided in Sections~\ref{sec:example_glm} and~\ref{sec:MLM_example}. 

Future enhancements of the package may aim to incorporate a broader spectrum of optimality criteria, such as A-optimality and E-optimality, as well as some models beyond GLMs and MLMs to expand its applicability.

\bibliography{CDsampling}




\end{article}

\begin{article}
\clearpage
\renewcommand{\baselinestretch}{1}

\markright{ \hbox{\footnotesize\rm The R Journal: Supplement
}\hfill\\[-13pt]
\hbox{\footnotesize\rm
}\hfill }

\markboth{\hfill{\footnotesize\rm Yifei Huang AND Liping Tong AND Jie Yang} \hfill}
{\hfill {\footnotesize\rm CDsampling} \hfill}

\renewcommand{\thefootnote}{}
$\ $\par \fontsize{12}{14pt plus.8pt minus .6pt}\selectfont


\centerline{\large\bf CDsampling: an R package for Constrained D-Optimal Sampling}
\vspace{2pt}
 \centerline{\large\bf in Paid Research Studies}
\vspace{.25cm}
 \centerline{Yifei Huang$^1$, Liping Tong$^2$, Jie Yang$^1$} 
\vspace{.4cm}
 \centerline{\it $^1$University of Illinois at Chicago, $^2$Advocate Aurora Health}
\vspace{.55cm}
 \centerline{\bf Supplementary Material}
\vspace{.55cm}
\fontsize{9}{11.5pt plus.8pt minus .6pt}\selectfont
\noindent

\noindent
{\bf S1 Examples of Fisher information matrix:} Two examples of finding Fisher information matrix with \CRANpkg{CDsampling}; \\
{\bf S2 Comparison of lift-one algorithm and constrained lift-one algorithm:} An example of comparison between lift-one algorithm and constrained lift-one algorithm;\\
{\bf S3 Comparison analysis of different sampling strategies:} Detailed R codes for a simulation study to compare various sampling strategies mentioned in the paper;\\
{\bf S4 Template functions of constraint index set $I$:} Detailed R codes for the example-specific functions of constraint index set $I$ in \CRANpkg{CDsampling};\\
\par

\setcounter{section}{0}
\setcounter{equation}{0}
\titleformat{\section}[hang]
  {\normalfont\large\bfseries}{\thesection}{1em}{}
\renewcommand\thesection{S\arabic{section}}

\numberwithin{equation}{section}

\setcounter{page}{1}

\fontsize{9}{12pt plus.8pt minus .6pt}\selectfont

\section{Examples of the Fisher information matrix}\label{sec:supp_fisher_example}
In Example~\ref{ex:GLM_Fisher} below, we provide an example of calculating the Fisher information matrix for GLM through the \CRANpkg{CDsampling} package. 

\begin{exmp}\label{ex:GLM_Fisher}
    Consider a research study under a logistic regression model 
    \begin{equation}\label{eq:logistic_two_covariate}
    \log\left(\frac{\mu_i}{1-\mu_i}\right) = \beta_0 + \beta_1 x_{i1} + \beta_2 x_{i2}
    \end{equation}
    where $\mu_i = E(Y_i\mid {\mathbf x}_i)$, $Y_i \in \{0,1\}$, ${\mathbf x}_i = (x_{i1}, x_{i2})^\top \in \{(-1, -1), (-1, 1), (1, -1)\}$ and parameters $\boldsymbol \beta = (\beta_0, \beta_1, \beta_2)^\top = (0.5, 0.5, 0.5)^\top$. The approximate allocation to the three possible design points, namely, ${\mathbf x}_i$'s, is $\mathbf w = (1/3,1/3,1/3)^\top$.
To calculate this Fisher information matrix $\mathbf F$ given the design ${\mathbf w}$, we may use \texttt{F\_func\_GLM()} function in the \CRANpkg{CDsampling} package. Additionally, \texttt{W\_func\_GLM()} function can be used to find the diagonal elements of the matrix $\mathbf W$ in the Fisher information matrix \eqref{eq:Fisher_GLM} of GLM. 

\begin{example}
> beta = c(0.5, 0.5, 0.5)
> X = matrix(data=c(1,-1,-1,1,-1,1,1,1,-1), byrow=TRUE, nrow=3) 
> w = c(1/3,1/3,1/3)
> F_func_GLM(w=w, beta=beta, X=X, link='logit')
Dimensions: 3 x 3 
Matrix:
------------------------------ 
     [,1]        [,2]        [,3]       
[1,]  0.23500371 -0.07833457 -0.07833457
[2,] -0.07833457  0.23500371 -0.07833457
[3,] -0.07833457 -0.07833457  0.23500371
------------------------------ 
\end{example} 
\hfill{$\Box$}
\end{exmp}
We also provide in Example~\ref{ex:MLM_Fisher} with the computation of the Fisher information matrix for MLM through the \CRANpkg{CDsampling} package. 

\begin{exmp}\label{ex:MLM_Fisher}
    We revisit the \texttt{trauma\_data} example previously analyzed in Section~\ref{sec:MLM_example}. Consider a research study under a cumulative logit npo model with $J=5$ response levels and covariates $(x_{i1}, x_{i2}) \in \{(1,0),(2,0),(3,0), (4,0),(1,1),(2,1),(3,1),$ $(4,1)\}$. The model can be written as the following four equations:
    \begin{eqnarray*}
    \log\left(\frac{\pi_{i1}}{\pi_{i2}+\dots+\pi_{i5}}\right) &=& \beta_{11}+\beta_{12}x_{i1}+\beta_{13}x_{i2}\\
    \log\left(\frac{\pi_{i1}+\pi_{i2}}{\pi_{i3}+\pi_{i4}+\pi_{i5}}\right) &=& \beta_{21}+\beta_{22}x_{i1}+\beta_{23}x_{i2}\\    \log\left(\frac{\pi_{i1}+\pi_{i2}+\pi_{i3}}{\pi_{i4}+\pi_{i5}}\right) &=& \beta_{31}+\beta_{32}x_{i1}+\beta_{33}x_{i2}\\
    \log\left(\frac{\pi_{i1}+\dots+\pi_{i4}}{\pi_{i5}}\right) &=&\beta_{41}+\beta_{42}x_{i1}+\beta_{43}x_{i2}
    \end{eqnarray*}
    where $i=1,\dots,8$. We assume that the true parameters $\boldsymbol \beta = (\hat\beta_{11},  \hat\beta_{12},  \hat\beta_{13}, \hat\beta_{21},  \hat\beta_{22},  \hat\beta_{23},  \hat\beta_{31}, $ $ \hat\beta_{32},  \hat\beta_{33},  \hat\beta_{41}, \hat\beta_{42},  \hat\beta_{43})^\top$ $=$ $(-4.047, -0.131, 4.214, -2.225, -0.376, 3.519, -0.302, -0.237,  2.420, 1.386,$ $ -0.120, 1.284)^\top$, and the approximate allocation for the eight design points is $\mathbf w = (1/8, \ldots, 1/8)^\top$. 

To calculate the Fisher information matrix $\mathbf F$ for the MLM in this example, we use the function \texttt{F\_func\_MLM()} in the \CRANpkg{CDsampling} package. In the R code below, $J$ is the number of levels of the response variable, $p$ is the number of parameters in the model, $m$ is the number of design points or subgroups, and $X_i$ is the model matrix of the $i$th design point. The function \texttt{F\_func\_MLM()} returns the Fisher information matrix ${\mathbf F}({\mathbf w})$ given ${\mathbf w}$.

\begin{example}
> J=5; p=12; m=8; 
> beta = c(-4.047, -0.131, 4.214, -2.225, -0.376, 3.519, -0.302, -0.237,  2.420, 1.386, 
-0.120,  1.284)
> Xi=rep(0,J*p*m); dim(Xi)=c(J,p,m)
> Xi[,,1] = rbind(c( 1, 1, 0, 0, 0, 0, 0, 0, 0, 0, 0, 0),
+                 c( 0, 0, 0, 1, 1, 0, 0, 0, 0, 0, 0, 0),
+                 c( 0, 0, 0, 0, 0, 0, 1, 1, 0, 0, 0, 0),
+                 c( 0, 0, 0, 0, 0, 0, 0, 0, 0, 1, 1, 0),
+                 c( 0, 0, 0, 0, 0, 0, 0, 0, 0, 0, 0, 0))
> Xi[,,2] = rbind(c( 1, 2, 0, 0, 0, 0, 0, 0, 0, 0, 0, 0),
+                 c( 0, 0, 0, 1, 2, 0, 0, 0, 0, 0, 0, 0),
+                 c( 0, 0, 0, 0, 0, 0, 1, 2, 0, 0, 0, 0),
+                 c( 0, 0, 0, 0, 0, 0, 0, 0, 0, 1, 2, 0),
+                 c( 0, 0, 0, 0, 0, 0, 0, 0, 0, 0, 0, 0))
> Xi[,,3] = rbind(c( 1, 3, 0, 0, 0, 0, 0, 0, 0, 0, 0, 0),
+                 c( 0, 0, 0, 1, 3, 0, 0, 0, 0, 0, 0, 0),
+                 c( 0, 0, 0, 0, 0, 0, 1, 3, 0, 0, 0, 0),
+                 c( 0, 0, 0, 0, 0, 0, 0, 0, 0, 1, 3, 0),
+                 c( 0, 0, 0, 0, 0, 0, 0, 0, 0, 0, 0, 0))
> Xi[,,4] = rbind(c( 1, 4, 0, 0, 0, 0, 0, 0, 0, 0, 0, 0),
+                 c( 0, 0, 0, 1, 4, 0, 0, 0, 0, 0, 0, 0),
+                 c( 0, 0, 0, 0, 0, 0, 1, 4, 0, 0, 0, 0),
+                 c( 0, 0, 0, 0, 0, 0, 0, 0, 0, 1, 4, 0),
+                 c( 0, 0, 0, 0, 0, 0, 0, 0, 0, 0, 0, 0))
> Xi[,,5] = rbind(c( 1, 1, 1, 0, 0, 0, 0, 0, 0, 0, 0, 0),
+                 c( 0, 0, 0, 1, 1, 1, 0, 0, 0, 0, 0, 0),
+                 c( 0, 0, 0, 0, 0, 0, 1, 1, 1, 0, 0, 0),
+                 c( 0, 0, 0, 0, 0, 0, 0, 0, 0, 1, 1, 1),
+                 c( 0, 0, 0, 0, 0, 0, 0, 0, 0, 0, 0, 0))
> Xi[,,6] = rbind(c( 1, 2, 1, 0, 0, 0, 0, 0, 0, 0, 0, 0),
+                 c( 0, 0, 0, 1, 2, 1, 0, 0, 0, 0, 0, 0),
+                 c( 0, 0, 0, 0, 0, 0, 1, 2, 1, 0, 0, 0),
+                 c( 0, 0, 0, 0, 0, 0, 0, 0, 0, 1, 2, 1),
+                 c( 0, 0, 0, 0, 0, 0, 0, 0, 0, 0, 0, 0))
> Xi[,,7] = rbind(c( 1, 3, 1, 0, 0, 0, 0, 0, 0, 0, 0, 0),
+                 c( 0, 0, 0, 1, 3, 1, 0, 0, 0, 0, 0, 0),
+                 c( 0, 0, 0, 0, 0, 0, 1, 3, 1, 0, 0, 0),
+                 c( 0, 0, 0, 0, 0, 0, 0, 0, 0, 1, 3, 1),
+                 c( 0, 0, 0, 0, 0, 0, 0, 0, 0, 0, 0, 0))
> Xi[,,8] = rbind(c( 1, 4, 1, 0, 0, 0, 0, 0, 0, 0, 0, 0),
+                 c( 0, 0, 0, 1, 4, 1, 0, 0, 0, 0, 0, 0),
+                 c( 0, 0, 0, 0, 0, 0, 1, 4, 1, 0, 0, 0),
+                 c( 0, 0, 0, 0, 0, 0, 0, 0, 0, 1, 4, 1),
+                 c( 0, 0, 0, 0, 0, 0, 0, 0, 0, 0, 0, 0))
> alloc = rep(1/8,m)
> F_func_MLM(w=rep(1/8,m), beta=beta, X=Xi, link='cumulative')
Dimensions: 12 x 12 
Matrix:
------------------------------------------------------------ 
      [,1]        [,2]        [,3]        [,4]        [,5]        [,6]       
 [1,]  0.44505694  1.37915564  0.43609135 -0.37247296 -1.21252053 -0.36379349
 [2,]  1.37915564  4.78410934  1.35694145 -1.21252053 -4.31436344 -1.19085831
 [3,]  0.43609135  1.35694145  0.43609135 -0.36379349 -1.19085831 -0.36379349
 [4,] -0.37247296 -1.21252053 -0.36379349  0.51192600  1.55177413  0.48018678
 [5,] -1.21252053 -4.31436344 -1.19085831  1.55177413  5.31193908  1.48241981
 [6,] -0.36379349 -1.19085831 -0.36379349  0.48018678  1.48241981  0.48018678
 [7,]  0.00000000  0.00000000  0.00000000 -0.09154268 -0.22027625 -0.07471168
 [8,]  0.00000000  0.00000000  0.00000000 -0.22027625 -0.64323991 -0.18445802
 [9,]  0.00000000  0.00000000  0.00000000 -0.07471168 -0.18445802 -0.07471168
[10,]  0.00000000  0.00000000  0.00000000  0.00000000  0.00000000  0.00000000
[11,]  0.00000000  0.00000000  0.00000000  0.00000000  0.00000000  0.00000000
[12,]  0.00000000  0.00000000  0.00000000  0.00000000  0.00000000  0.00000000
      [,7]        [,8]        [,9]        [,10]       [,11]       [,12]      
 [1,]  0.00000000  0.00000000  0.00000000  0.00000000  0.00000000  0.00000000
 [2,]  0.00000000  0.00000000  0.00000000  0.00000000  0.00000000  0.00000000
 [3,]  0.00000000  0.00000000  0.00000000  0.00000000  0.00000000  0.00000000
 [4,] -0.09154268 -0.22027625 -0.07471168  0.00000000  0.00000000  0.00000000
 [5,] -0.22027625 -0.64323991 -0.18445802  0.00000000  0.00000000  0.00000000
 [6,] -0.07471168 -0.18445802 -0.07471168  0.00000000  0.00000000  0.00000000
 [7,]  0.29320484  0.71978889  0.16205254 -0.10435894 -0.25399393 -0.06194131
 [8,]  0.71978889  2.13312603  0.41294396 -0.25399393 -0.74842743 -0.15305771
 [9,]  0.16205254  0.41294396  0.16205254 -0.06194131 -0.15305771 -0.06194131
[10,] -0.10435894 -0.25399393 -0.06194131  0.17861575  0.45287619  0.06925715
[11,] -0.25399393 -0.74842743 -0.15305771  0.45287619  1.37180187  0.17395849
[12,] -0.06194131 -0.15305771 -0.06194131  0.06925715  0.17395849  0.06925715
------------------------------------------------------------ 
\end{example}

\hfill{$\Box$}
\end{exmp}

\section{Comparison of lift-one algorithm and constrained lift-one algorithm}\label{sec:supp_comparison}
\begin{exmp}\label{ex:simple_logistic}
    We consider the same logistic regression model \eqref{eq:logistic_two_covariate} as in Example~\ref{ex:GLM_Fisher}. That is,
$$\log(\mu_i/(1-\mu_i)) = \beta_0 + \beta_1 x_{i1} + \beta_2 x_{i2}$$ with ${\mathbf x}_i = (x_{i1}, x_{i2})^\top \in \{(-1, -1), (-1, 1), (1, -1)\}$ and $(\beta_0, \beta_1, \beta_2) = (0.5, 0.5, 0.5)$. 
To calculate the ${\mathbf W}$  matrix in \eqref{eq:Fisher_GLM} for this model, we use the following R codes: 

\begin{example}
> beta = c(0.5, 0.5, 0.5) 
> X = matrix(data=c(1,-1,-1,1,-1,1,1,1,-1), byrow=TRUE, nrow=3)
> W_matrix = W_func_GLM(X=X, b=beta)     
\end{example}

If we don't consider any additional constraints on allocations, we may run the unconstrained lift-one algorithm using \texttt{liftone\_GLM()} for GLMs or \texttt{liftone\_MLM()} for MLMs. 

\begin{example}
> w00 = c(1/6, 1/6, 2/3) #an arbitrary starting allocation
> liftone_GLM(X=X, W=W_matrix, reltol=1e-10, maxit=100, random=FALSE, nram=3, w00=w00)   

Optimal Sampling Results:
================================================================================
Optimal approximate allocation:
   1      2      3     
w  0.3333 0.3333 0.3333
w0 0.1667 0.1667 0.6667
--------------------------------------------------------------------------------
Maximum :
0.0077 
--------------------------------------------------------------------------------
itmax :
7.0 
--------------------------------------------------------------------------------
convergence :
TRUE 
--------------------------------------------------------------------------------
\end{example}

This suggests that in the feasible allocation set $S_0= \{(w_1, w_2, w_3)^\top$ $\in$ $\mathbb{R}^3 \mid w_i \geq 0, i=1, \ldots, 3; \sum_{i=1}^3 w_i = 1\}$, the D-optimal allocation is $\mathbf w = (w_1, w_2, w_3)=(1/3, 1/3, 1/3)$. 
\hfill{$\Box$}
\end{exmp}

\begin{exmp}\label{ex:simple_logistic_r1_r2}
We consider the same logistic regression model \eqref{eq:logistic_two_covariate} as in Examples~\ref{ex:GLM_Fisher} and \ref{ex:simple_logistic}. In this example, we aim for the constrained D-optimal allocation in $S = \{(w_1, w_2, w_3)^\top \in S_0 \mid w_1 \leq \frac{1}{6}, w_3 \geq \frac{8}{15}, 4w_1 \geq w_3\}$. 
The additional constraints on $S_0$ 
can be defined using the following R codes:  
\begin{example}
> g.con = matrix(,nrow=10, ncol=3)
> g.con[1,]=c(1,1,1)
> g.con[2:7, ]=rbind(diag(3),diag(3))
> g.con[8,]=c(1,0,0)
> g.con[9,]=c(0,0,1)
> g.con[10,]=c(4,0,-1)
> g.dir = c("==", rep(">=",3), rep("<=",3),"<=", ">=", ">=")
> g.rhs = c(1, rep(0,3), rep(1,3), 1/6, 8/15, 0)
\end{example}

We also need to find the upper boundary and lower boundary functions for $[r_{i1}, r_{i2}]$ in step $3^\circ$ of the constrained lift-one algorithm to meet the conditions in $S$.
The $r_{i1}$ and $r_{i2}$ can be obtained as follows: 

\noindent
{\it Case one:} If $i =1$, then ${\mathbf w}_i(z) = (z, \frac{1-z}{1-w_1}w_2, \frac{1-z}{1-w_1}w_3)^\top \in S$ if and only if 
\begin{equation*}
\left\{\begin{array}{cl}
0 \leq z \leq 1/6 \\
0 \leq \frac{1-z}{1-w_1}w_2 \leq 1 \\
8/15 \leq \frac{1-z}{1-w_1} w_3 \leq 1 \\
4z \ge \frac{1-z}{1-w_1}w_3
\end{array}\right.
\end{equation*}
which is equivalent to \begin{equation*}
\left\{\begin{array}{cl}
0 \leq z \leq 1/6 \\
1-\frac{1-w_1}{w_2} \leq z \leq 1\\
1-\frac{1-w_1}{w_3}\leq z \leq 1-\frac{8(1-w_1)}{15w_3}\\
z \geq \frac{w_3}{4-4w_1+w_3}
\end{array}\right.
\end{equation*}
Therefore, 
\begin{equation*}
 \left\{\begin{array}{cl}
 r_{i1} = & \max\{1-\frac{1-w_1}{w_2}, 1-\frac{1-w_1}{w_3}, \frac{w_3}{4-4w_1+w_3}\}\\
r_{i2}  = & \min\{1/6, 1-\frac{8(1-w_1)}{15w_3}\}   
 \end{array}\right.
 \end{equation*}

\noindent
{\it Case two:} If $i=2$, then ${\mathbf w}_i(z)=(\frac{1-z}{1-w_2}w_1, z, \frac{1-z}{1-w_2}w_3)^\top \in S$ if and only if 
\begin{equation*}
\left\{\begin{array}{cl}
0 \leq \frac{1-z}{1-w_2}w_1 \leq 1/6 \\
0 \leq z \leq 1 \\
8/15 \leq \frac{1-z}{1-w_2}w_3 \leq 1\\
4\frac{1-z}{1-w_2}w_1 \ge \frac{1-z}{1-w_2}w_3 
\end{array}\right.
\end{equation*}
Similarly, we obtain
\begin{equation*}
 \left\{\begin{array}{cl}
 r_{i1} = & \max\{0, 1-\frac{1-w_2}{6w_1}, 1-\frac{1-w_2}{w_3} \}\\
r_{i2}  = & 1-\frac{8(1-w_2)}{15w_3}  
 \end{array}\right.
 \end{equation*}

 \noindent
{\it Case three:} If $i=3$, then ${\mathbf w}_i(z)=(\frac{1-z}{1-w_3}w_1, \frac{1-z}{1-w_3}w_2, z)^\top \in S$ if and only if 
\begin{equation*}
\left\{\begin{array}{cl}
0 \leq \frac{1-z}{1-w_3}w_1 \leq 1/6 \\
0 \leq \frac{1-z}{1-w_3}w_2 \leq 1 \\
8/15 \leq z \leq 1\\
4\frac{1-z}{1-w_3}w_1 \ge z 
\end{array}\right.
\end{equation*}
Similarly, we obtain
\begin{equation*}
 \left\{\begin{array}{cl}
 r_{i1} = & \max\{8/15, 1-\frac{1-w_3}{6w_1}, 1-\frac{1-w_3}{w_2}\}\\
r_{i2}  = & \frac{4w_1}{1+4w_1-w_3}   
 \end{array}\right.
 \end{equation*}

We may code $r_{i1}, r_{i2}$ as functions in R as follows: 

\begin{example}
> lower.bound = function(i, w){
+   if(i == 1){
+     return(max(1-(1-w[i])/w[2], 1-(1-w[i])/w[3], (w[3])/(4-4*w[i]+w[3])))}
+   if(i == 2){
+     return(max(0, 1-((1-w[i])/(6*w[1])), 1-(1-w[i])/w[3]))}
+   if(i == 3){
+     return(max(8/15, (1-(1-w[i])/(6*w[1])), 1-(1-w[i])/w[2]))}}

> upper.bound = function(i, w){
+   if(i == 1){
+     return(min(1/6, 1-(8*(1-w[i])/(15*w[3]))))}
+   if(i == 2){
+     return(1-(8*(1-w[i])/(15*w[3])))}
+   if(i == 3){
+     return((4*w[1])/(1+4*w[1]-w[i]))}}
\end{example}

Now we are ready to find the constrained D-optimal allocation in $S$ using the constrained lift-one function \texttt{liftone\_constrained\_GLM()}.

\begin{example}
> set.seed(123)
> liftone_constrained_GLM(X=X, W=W_matrix, g.con=g.con, g.dir=g.dir, g.rhs=g.rhs, 
+ lower.bound=lower.bound, upper.bound=upper.bound, reltol=1e-10, maxit=100, 
+ random=FALSE, nram=3, w00=w00, epsilon=1e-8) 

Optimal Sampling Results:
================================================================================
Optimal approximate allocation:
   1      2      3     
w  0.1667 0.3    0.5333
w0 0.1667 0.1667 0.6667
--------------------------------------------------------------------------------
maximum :
0.0055 
--------------------------------------------------------------------------------
convergence :
TRUE 
--------------------------------------------------------------------------------
itmax :
1.0 
--------------------------------------------------------------------------------
deriv.ans :
0.0199, 0.0026, -0.0133 
--------------------------------------------------------------------------------
gmax :
0.0 
--------------------------------------------------------------------------------
reason :
"gmax <= 0" 
--------------------------------------------------------------------------------
\end{example}

The result indicates that with the feasible allocation set $S$, our constrained D-optimal allocation converge to a different point with the D-optimal approximate allocation $(w_1, w_2, w_3)= (\frac{1}{6}, \frac{3}{10}, \frac{8}{15})$ and the lift-one loop breaks because $\max g({\mathbf w}) \leq 0$, where $ g({\mathbf w})= \sum_{i=1}^m w_i (1-w_i^*) f_i'(w_i^*)$ in step $8^\circ$ of the constrained lift-one algorithm (Figure~\ref{fig:constrained_liftone_algo}). 
\hfill{$\Box$}
\end{exmp}

\section{Comparison analysis of different sampling strategies}\label{sec:supp_simulation}
In this section, we present the detailed simulation codes for comparing the simple random sample without replacement (SRSWOR), constrained D-optimal allocation, constrained EW D-optimal allocation, and constrained uniform allocation as discussed in Figure~\ref{fig:RMSE_GLM} of Section~\ref{sec:example_glm}.

We begin by defining the model matrix, coefficients, and constraints, consistent with the specifications in Section~\ref{sec:example_glm}, using the following codes:

\begin{example}
> beta = c(0, 3, 3, 3)
> X=matrix(data=c(1,0,0,0,1,0,1,0,1,0,0,1,1,1,0,0,1,1,1,0,1,1,0,1), ncol=4, byrow=TRUE)
> W=CDsampling::W_func_GLM(X=X, b=beta, link="logit")
> rc = c(50, 40, 10, 200, 150, 50)/200 #available volunteers/sample size
> m = 6
> g.con = matrix(0,nrow=(2*m+1), ncol=m)
> g.con[1,] = rep(1, m)
> g.con[2:(m+1),] = diag(m)
> g.con[(m+2):(2*m+1), ] = diag(m)
> g.dir = c("==", rep("<=", m), rep(">=", m))
> g.rhs = c(1, rc, rep(0, m))

> lower.bound=function(i, w){
+   nsample = 200
+   rc = c(50, 40, 10, 200, 150, 50)/nsample
+   m=length(w) #num of categories
+   temp = rep(0,m)
+   temp[w>0]=1-pmin(1,rc[w>0])*(1-w[i])/w[w>0];
+   temp[i]=0;
+   max(0,temp);
+ }

> upper.bound=function(i, w){
+   nsample = 200
+   rc = c(50, 40, 10, 200, 150, 50)/nsample
+   m=length(w) #num of categories
+   rc[i];
+   min(1,rc[i]);
+ }

\end{example}

To find the constrained D-optimal allocation, we use the following codes: 
\begin{example}
> set.seed(123)
> approximate_design = liftone_constrained_GLM(X=X, W=W, g.con=g.con, 
+ g.dir=g.dir, g.rhs=g.rhs, lower.bound=lower.bound, upper.bound=upper.bound, 
+ reltol=1e-10, maxit=100, random=TRUE, nram=4, w00=NULL, epsilon=1e-8)

> exact_design = approxtoexact_constrained_func(n=200, w=approximate_design$w, 
+ m=6, beta=beta, link='logit', X=X, Fdet_func=Fdet_func_GLM, 
+ iset_func=iset_func_trial)
\end{example}

To determine the EW D-optimal allocation with the independent uniform priors $\beta_0\sim$ uniform$(-2,2)$, $\beta_1\sim$ uniform$(-1,5)$, $\beta_{21}\sim$ uniform$(-1,5)$, and $\beta_{22}\sim$ uniform(-1,5), as outlined in Section~\ref{sec:example_glm}, we utilize the same codes provided in Section~\ref{sec:example_glm}:
\begin{example}
> library(cubature)
> unif.prior <- rbind(c(-2, -1, -1, -1), c(2,  5,  5, 5))
> W.EW.unif = matrix(rep(0,6))
> for (i in 1:6){
+   x = matrix((cbind(1, unique(X)))[i,])
+   W.EW.unif[i] = hcubature(function(beta) dunif(beta[1], min=unif.prior[1,1], 
max=unif.prior[2,1])*dunif(beta[2], min=unif.prior[1,2], 
max=unif.prior[2,2])*dunif(beta[3], min=unif.prior[1,3], 
max=unif.prior[2,3])*dunif(beta[4], min=unif.prior[1,4], max=unif.prior[2,4])*
(exp(x[1]*beta[1]+x[2]*beta[2]+x[3]*beta[3]+x[4]*beta[4])/(1+exp(x[1]*beta[1]+
x[2]*beta[2]+x[3]*beta[3]+x[4]*beta[4]))^2), lowerLimit = unif.prior[1,], 
upperLimit  = unif.prior[2,])$integral
+ }

> set.seed(602)
> approximate_design_EW = liftone_constrained_GLM(X=X, W=W.EW.unif,  g.con=g.con, 
+ g.dir=g.dir, g.rhs=g.rhs, lower.bound=lower.bound, upper.bound=upper.bound, 
+ reltol=1e-12, maxit=100, random=TRUE, nram=12, w00=NULL, epsilon=1e-12)

> exact_design_EW = approxtoexact_constrained_func(n=200, w=approximate_design_EW$w,
+ m=6, beta=beta, link='logit', X=X, Fdet_func=Fdet_func_GLM, iset_func=iset_func_trial)
\end{example}

To find the constrained uniform allocation, we use the following codes: 
\begin{example}
> w00 = rep(1/200, 6)
> unif_design = approxtoexact_constrained_func(n=200, w=w00, m=6, beta=NULL, link=NULL,
+ X=NULL, Fdet_func=Fdet_func_unif, iset_func=iset_func_trial)
\end{example}

Next, we conduct 100 simulations and record the corresponding Root Mean Square Error (RMSE) from both the full data ($500$ patients) fitted model and the models fitted using the aforementioned samplings with $200$ patients. The R codes used for this process are as follows:

\begin{example}
> nsimu=100
> p=3 # 3 covariates (excluding intercept)
> set.seed(666)
> seeds <- sample(1:10000, nsimu)   #random seeds

> # matrix for recording beta estimates
> beta.estimate <- rep(NA, nsimu*(p+1)*(5)) 
> dim(beta.estimate) = c(nsimu, p+1, 5) # [1] 100   4   8
> dimnames(beta.estimate)[[2]] = paste0("beta",0:p,seq="")
> dimnames(beta.estimate)[[3]] <- c('full', "SRSWOR", "Unif", "local_Dopt", "EW_Unif")

> # generate X for original 500 patients
> # gender group (0 for female and 1 for male)
> gender = c(rep(0,50),rep(0,40),rep(0,10),rep(1,200),rep(1,150),rep(1,50)) 
> # age group (0 for 18~25, 1 for 26~64, and 2 for 65 or above)
> age1 = c(rep(0,50), rep(1,40), rep(0,10), rep(0,200), rep(1,150), rep(0,50)) 
> age2 = c(rep(0,50), rep(0,40), rep(1,10), rep(0,200), rep(0,150), rep(1,50))
> X.original = cbind(gender, age1, age2)
> #probability of Y = 1 given X using coefficient beta under logistic model
> prob = exp(beta 

> label = rep(0, length(X.original[,1])) #create label to store category of the population data
> for(k in 1:length(X.original[,1])){
+   for (m in 1:m){
+     if((X.original[k,]==unique(X.original)[m,])[1] & (X.original[k,]==unique(X.original)[m,])[2] 
& (X.original[k,]==unique(X.original)[m,])[3]){
+       label[k] = m
+     }
+   }
+ }
> n = tabulate(label) #number of subjects available in each category
> sampling_func = function(label, n.sample, X, s, n, m){
+   J = 0                                  
+   j = rep(0, m)                   
+   c = rep(0, m)                   
+   isample = vector()                     
+   while(J < n.sample){
+     for(k in 1:length(X[,1])){
+       i = label[k]
+       c[i] = c[i] + 1
+       if(j[i]<s[i]){
+         if(runif(1) < (s[i] - j[i])/(n[i] - c[i] + 1)){
+           j[i]=j[i]+1
+           J = J + 1
+           isample = c(isample, k)
+         }
+       }
+     }
+   }
+   return(isample)
+ }

> s_unif = unif_design$allocation #num to sample in constrained uniform from each category
> s_localD = exact_design$allocation #num to sample in local D-optimal from each category
> s_EWD = exact_design_EW$allocation #num to sample in EW D-optimal from each category

> for(isimu in 1:nsimu){
+   set.seed(seeds[isimu])
+   ##Generate Y response under logistic model
+   Y = rbinom(n=length(X.original[,1]), size=1, prob = prob) 

+   #Estimate beta with full data
+   beta.estimate[isimu,,"full"] <- glm(Y~X.original,family = "binomial")$coefficients

+   #SRSWOR sample
+   isample1 <- sample(x=1:500, size=200); 
+   X1=X.original[isample1,];
+   Y1=Y[isample1];

+   #Estimate beta with SRSWOR sampling
+   beta.estimate[isimu,,"SRSWOR"] <- glm(Y1~X1,family = "binomial")$coefficients

+   #Uniform sample
+   #stratified function returns s[k] (num of samples plan to collect for ith category)
+   isample2 = sampling_func(label=label, n.sample=200, X=X.original, s=s_unif, n=n, m=m)
+   X2=X.original[isample2,];
+   Y2=Y[isample2];

+   #Estimate beta with Unif sampling
+   beta.estimate[isimu,,"Unif"] <- glm(Y2~X2,family = "binomial")$coefficients

+   #Local D-opt Sample
+   isample3 = sampling_func(label=label, n.sample=200, X=X.original, s=s_localD, n=n,m=m)
+   X3=X.original[isample3,];
+   Y3=Y[isample3];

+   #Estimate beta with local D-optimal sampling
+   beta.estimate[isimu,,"local_Dopt"] <- glm(Y3~X3,family = "binomial")$coefficients

+   #EW D-opt Sample
+   isample4 = sampling_func(label=label, n.sample=200, X=X.original, s=s_EWD, n=n, m=m)
+   X4=X.original[isample4,];
+   Y4=Y[isample4];
+   #Estimate beta with EW D-optimal sampling
+   beta.estimate[isimu,,"EW_Unif"] <- glm(Y4~X4,family = "binomial")$coefficients
+ }
\end{example}

We prepare the dataset for figure creation using the following R codes:
\begin{example}
> beta.noint = beta[2:(p+1)]
> ## Estimating beta0 (intercept) compared to true
> btemp0 = abs(beta.estimate[,1,]) #abs(beta0_est - beta0), beta0 = 0
> mse.b0.mean=apply(btemp0,02,mean, na.rm = TRUE)
> mse.b0.sd=apply(btemp0,2,sd) #sd of beta0 mse

> ## Estimating beta1-beta4 compared to true
> btemp=1/3*(beta.estimate[,2:(p+1),]-beta.noint)^2 #square of beta's error
> btemp=sqrt(apply(btemp,c(1,3),sum, na.rm = TRUE)) 
> mse.b5all.mean=apply(btemp,2,mean, na.rm = TRUE) #apply mean over column
> mse.b5all.sd=apply(btemp,2,sd, na.rm = TRUE)

> ## separated mse sd for each beta parameter
> btemp1 = sqrt((beta.estimate[,2,] - beta[2])^2)
> mse.b1.mean=apply(btemp1,2,mean, na.rm = TRUE) 
> mse.b1.sd=apply(btemp1,2,sd, na.rm = TRUE) #sd of mse

> btemp2 = sqrt((beta.estimate[,3,] - beta[3])^2)
> mse.b2.mean=apply(btemp2,2,mean, na.rm = TRUE) 
> mse.b2.sd=apply(btemp2,2,sd, na.rm = TRUE) #sd of mse

> btemp3 = sqrt((beta.estimate[,4,] - beta[4])^2)
> mse.b3.mean=apply(btemp3,2,mean, na.rm = TRUE) 
> mse.b3.sd=apply(btemp3,2,sd, na.rm = TRUE) #sd of mse

> output_simulation_rmse = as.data.frame(rbind(cbind(btemp0, category="beta0"), 
cbind(btemp, category="all_coef"), cbind(btemp1, category="beta1"), cbind(btemp2, 
category="beta21"), cbind(btemp3, category="beta22")))

> library(data.table)
> output_simulation_rmse.long=melt(setDT(output_simulation_rmse), id.vars=c("category"), 
variable.name='Method') 
> colnames(output_simulation_rmse.long)[3]='RMSE'
> colnames(output_simulation_rmse.long)[1]='Coef'
\end{example}

Finally, Figure~\ref{fig:RMSE_GLM} is generated using the following R codes: 
\begin{example}
> library(ggplot2)
> library(dplyr)
> output_simulation_rmse.long 
+ ggplot(aes(x=Coef, y=as.numeric(RMSE), fill=factor(Method))) + 
+ geom_boxplot(alpha=0.7) + 
+ stat_summary(fun=mean, color = "black", position = position_dodge(0.75),
+ geom = "point", shape = 18, size = 2, show.legend = FALSE)+ 
+ scale_fill_grey(start = 0, end = 1, name="Design Method") +
+ xlab("Coefficient")+ylab("RMSE")+ theme_bw()+
+ facet_wrap(~Coef, scales='free')
\end{example}

\section{Template functions of constraint index set $I$}\label{sec:supp_iset_codes}

\begin{example}
> iset_func_trauma <- function(allocation){
+   iset = rep(TRUE,8)
+   if(sum(allocation[1:4])>=392){iset[1:4]=FALSE}
+   if(sum(allocation[5:8])>=410){iset[5:8]=FALSE}
+   return(iset)
+ }turn(iset)
}
\end{example}

\begin{example}
> iset_func_trial <- function(allocation){
+   Ni = c(50, 40, 10, 200, 150, 50)
+   return(allocation < Ni)
+ } 
\end{example}

\end{article}

\end{document}